\documentclass[12pt]{article}
\pdfoutput=1
\usepackage{epsfig}
\usepackage{tabularx}
      \def\di{\displaystyle}
      
      \def\bS{{\bf S}}
      \def\bl{{\bf l}}
      \def\bp{{\bf p}}
      \def\bq{{\bf q}}
      \def\br{{\bf r}}
      \def\bs{{\bf s}}
      
      \def\A{{\cal A}}
      
      \def\E{{\cal E}}
      \def\F{{\cal F}}
      \def\H{{\cal H}}
      
      \def\K{{\cal K}}
      \def\L{{\cal L}}
      
      \def\P{{\cal P}}
      \def\R{{\cal R}}

      \def\u{{{\uparrow\downarrow}}}
      \def\d{{{\downarrow\uparrow}}}
\begin{document}

\begin{center}
{\large\bf 
 ORBITAL AND SPIN SCISSORS MODES IN SUPERFLUID NUCLEI
}\\
\vspace*{4mm}
{\large E.B. Balbutsev, I.V. Molodtsova}\\
{\it Joint Institute for Nuclear Research, 141980 Dubna, Moscow Region,
Russia}\\
\vspace*{1mm}
{\large P. Schuck}\\
{\it Institut de Physique Nucl\'eaire, IN2P3-CNRS, Universit\'e Paris-Sud,\\
F-91406 Orsay C\'edex, France;\\
Laboratoire de Physique et Mod\'elisation des Milieux Condens\'es,
CNRS and Universit\'e Joseph Fourier,
25 avenue des Martyrs BP166, F-38042 Grenoble C\'edex 9, France}
\end{center}

\begin{abstract}
Nuclear scissors modes are considered in the frame of Wigner function moments method
generalized to take into account spin degrees of freedom and pair correlations simultaneously.
A new source of nuclear magnetism, connected with
counter-rotation of spins up and down around the symmetry axis
(hidden angular momenta), 
is discovered. Its inclusion into
the theory allows one to improve substantially the agreement with experimental data in the 
description of energies and transition probabilities of scissors modes in
rare earth nuclei.
\end{abstract}



\section{Introduction}

The nuclear scissors mode was predicted \cite{Hilt}--\cite{Lo} 
as a counter-rotation of protons against neutrons in deformed nuclei.  
However, its collectivity turned out to be small. From RPA results which
 were in qualitative agreement with experiment, it was even questioned 
 whether this mode is collective at all \cite{Zaw,Sushkov}. 
Purely phenomenological models (such as, e.g., 
the two rotors model \cite{Lo2000}) and the sum rule approach~\cite{Lipp} 
did not clear up the situation in this 
respect. Finally in a recent review \cite{Heyd} it is concluded 
that the scissors mode is "weakly collective, but strong 
on the single-particle scale" and further: "The weakly
collective scissors mode excitation has become an ideal test of models
-- especially microscopic models -- of nuclear vibrations. Most models
are usually calibrated to reproduce properties of strongly collective
excitations (e.g. of $J^{\pi}=2^+$ or $3^-$ states, giant resonances,
...). Weakly-collective phenomena, however, force the models to make
genuine predictions and the fact that the transitions in question are
strong on the single-particle scale makes it impossible to dismiss
failures as a mere detail, especially in the light of the overwhelming
experimental evidence for them in many nuclei \cite{Kneis,Richt}."

The Wigner Function
Moments (WFM) or phase space moments method turns out to be very 
useful in this
situation. On the one hand it is a purely microscopic method, because
it is based on the Time Dependent Hartree-Fock (TDHF) equation. On the
other hand the method works with average values (moments) of operators
which have a direct relation to the considered phenomenon and, thus, make a 
natural bridge with the macroscopic description. This 
makes it an ideal instrument to describe the basic characteristics 
(energies and excitation probabilities) of collective excitations such as,
in particular, the scissors mode.

Further developments of the WFM
method, namely, the switch from TDHF
to TDHF-Bogoliubov (TDHFB) equations, i.e. taking into account pair correlations, allowed
us to improve considerably the quantitative description of the 
scissors mode \cite{Malov,Urban}: for rare earth nuclei the energies were
reproduced with
$\sim 10\%$ accuracy and B(M1) values were reduced by about a factor of two  
with respect to their non superfluid values. 
However, they remained about two times too high with respect to experiment.
We have suspected, that the reason of this last discrepancy is hidden in the spin 
degrees of freedom, which were so far ignored by the WFM method.

\begin{figure}[h]
\centering\includegraphics[width=5cm]{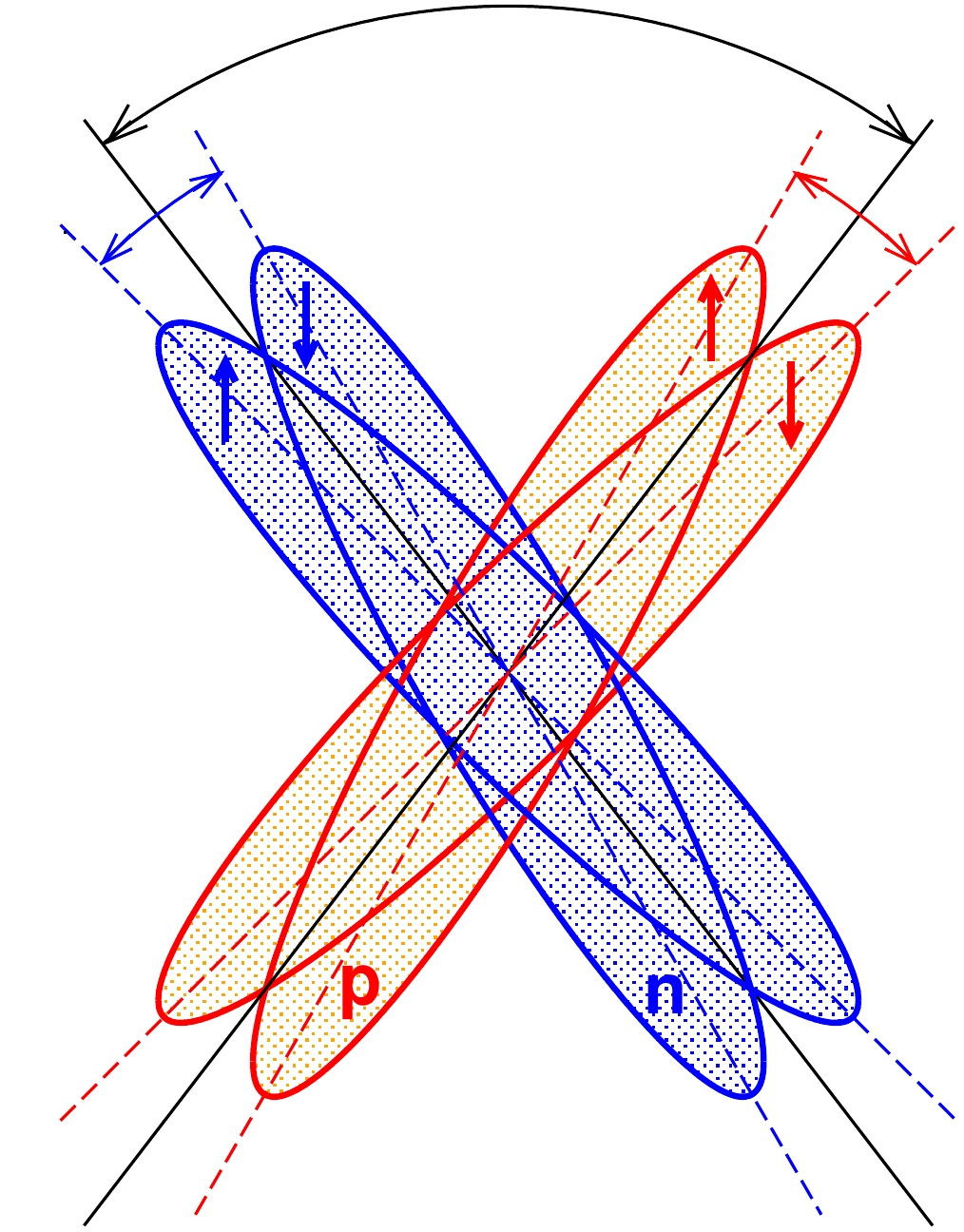}
\caption{Pictorial representation of two intermingled scissors: the orbital (neutrons 
versus protons) scissors + spin (spin-up nucleons versus spin-down nucleons)
scissors. Arrows inside of ellipses show the direction of spin projections.
 {\bf p} - protons, {\bf n} - neutrons.}
\label{fig0}\end{figure}

 In a recent paper \cite{BaMo} the WFM method was 
applied for the first time to solve the TDHF equations including spin
dynamics.
As a first step, only the spin orbit interaction was included in the
consideration, as the most important
one among all possible spin dependent interactions because it enters 
into the mean field. 
 The most remarkable result was the discovery of a new type
of nuclear collective motion: rotational oscillations of "spin-up"
nucleons with respect of "spin-down" nucleons (the spin scissors mode). 
It turns out that the experimentally 
observed group of peaks in the energy interval 2-4 MeV corresponds 
very likely to
two different types of motion: the orbital scissors mode and this new kind 
of mode, i.e. the spin scissors mode. The pictorial view of these two intermingled scissors
is shown on Fig.~\ref{fig0}, which is just the modification (or generalization) of the 
classical picture for 
the orbital scissors (see, for example, \cite{Lo2000,Heyd}).


The next step was done in the paper \cite{BaMoPRC}, where the influence of the spin-spin
interaction on the scissors modes was studied. There was hope
that, due to spin dependent interactions, some part of the 
force of M1 transitions will be shifted to the energy region of 5-10 MeV (the area of a
spin-flip resonance),
decreasing in such a way the M1 force of scissors. However, these expectations were not realised. It turned out that the spin-spin interaction does not change the general
picture of the positions of excitations described in \cite{BaMo} pushing all levels
up proportionally to its strength without changing their order. The most interesting
result concerns the B(M1) values of both scissors
-- the spin-spin interaction strongly redistributes M1 strength in  favour
of the spin scissors mode practically without changing their summed strength.

In the present work we suggest a generalization of the WFM method which takes into account
spin degrees of freedom and pair correlations simultaneously. According to our previous 
calculations these two factors, working together, should improve considerably the agreement 
between the theory and experiment in the description of nuclear scissors modes.

The paper is organized as follows.
 In Sec. 2 the TDHFB equations for the 2x2 normal and anomalous density matrices are
formulated and their Wigner transform is found.
In Sec. 3 the model Hamiltonian and the mean field are analyzed.
In Sec. 4 the collective variables are defined and the respective 
dynamical equations are derived.
In Sec. 5 the choice of parameters and the results of calculations of energies and B(M1)
values of two scissors modes are discussed.
The phenomenon of counter-rotating angular momenta with spin up/down,
which can be considered also as a phenomenon of hidden angular momenta, is analysed in Sec. 6.
Results of calculations for 26 nuclei in the rare earth region 
are discussed in Sec. 7.
The summary of main results is given in the conclusion section. 
The mathematical details are concentrated in Appendices A, B, C, D.

\section{Wigner transformation of TDHFB equations}

\hspace{5mm} The Time-Dependent Hartree--Fock--Bogoliubov (TDHFB)
equations in matrix formulation are
\cite{Solov,Ring}
\begin{equation}
i\hbar\dot\R=[\H,\R]
\label{tHFB}
\end{equation}
with
\begin{equation}
\R={\hat\rho\qquad-\hat\kappa\choose-\hat\kappa^{\dagger}\;\;1-\hat\rho^*},
\quad\H={\hat
h\quad\;\;\hat\Delta\choose\hat\Delta^{\dagger}\quad-\hat h^*}
\end{equation}
The normal density matrix $\hat \rho$ and Hamiltonian $\hat h$ are
hermitian whereas the abnormal density $\hat \kappa$ and the pairing
gap $\hat \Delta$ are skew symmetric: $\hat \kappa^{\dagger}=-\hat
\kappa^*$, $\hat \Delta^{\dagger}=-\hat \Delta^*$.

The detailed form of the TDHFB equations is
\begin{eqnarray}
&& i\hbar\dot{\hat\rho} =\hat h\hat\rho -\hat\rho\hat h
-\hat\Delta \hat\kappa ^{\dagger}+\hat\kappa \hat\Delta^\dagger,
\nonumber\\
&&-i\hbar\dot{\hat\rho}^*=\hat h^*\hat\rho ^*-\hat\rho ^*\hat h^*
-\hat\Delta^\dagger\hat\kappa +\hat\kappa^\dagger\hat\Delta ,
\nonumber\\
&&-i\hbar\dot{\hat\kappa} =-\hat h\hat\kappa -\hat\kappa \hat h^*+\hat\Delta
-\hat\Delta \hat\rho ^*-\hat\rho \hat\Delta ,
\nonumber\\
&&-i\hbar\dot{\hat\kappa}^\dagger=\hat h^*\hat\kappa^\dagger
+\hat\kappa^\dagger\hat h-\hat\Delta^\dagger
+\hat\Delta^\dagger\hat\rho +\hat\rho^*\hat\Delta^\dagger .
\label{HFB}
\end{eqnarray}
It is easy to see that the second and fourth equations are complex 
conjugate to the first and third ones respectively.
Let us consider their matrix form in coordinate 
space keeping all spin indices $s, s', s''$:
\begin{eqnarray}
i\hbar\langle \br,s|\dot{\hat\rho}|\br'',s''\rangle  =
\hspace{12cm}
\nonumber\\
\hspace{10mm}
\sum_{s'}\int\!d^3r'\left(
\langle \br,s|\hat h|\br',s'\rangle \langle \br',s'|\hat\rho|\br'',s''\rangle  
-\langle \br,s|\hat\rho|\br',s'\rangle \langle \br',s'|\hat h|\br'',s''\rangle \right.
\nonumber\\
\hspace{10mm}
\left.
-\langle \br,s|\hat\Delta|\br',s'\rangle \langle \br',s'|\hat\kappa^{\dagger}|\br'',s''\rangle  
+\langle \br,s|\hat\kappa|\br',s'\rangle \langle \br',s'|\hat\Delta^{\dagger}|\br'',s''\rangle 
\right),
\nonumber\\
i\hbar\langle \br,s|\dot{\hat\kappa}|\br'',s''\rangle  = -\langle \br,s|\hat\Delta|\br'',s''\rangle 
\hspace{85mm}
\nonumber\\
\hspace{10mm}
+\sum_{s'}\int\!d^3r'\left(
\langle \br,s|\hat h|\br',s'\rangle \langle \br',s'|\hat\kappa|\br'',s''\rangle  
+\langle \br,s|\hat\kappa|\br',s'\rangle \langle \br',s'|\hat h^*|\br'',s''\rangle \right.
\nonumber\\
\hspace{10mm}
\left.
+\langle \br,s|\hat\Delta|\br',s'\rangle \langle \br',s'|\hat\rho^*|\br'',s''\rangle  
+\langle \br,s|\hat\rho|\br',s'\rangle \langle \br',s'|\hat\Delta|\br'',s''\rangle 
\right),
\nonumber\\
i\hbar\langle \br,s|\dot{\hat\rho}^*|\br'',s''\rangle  =
\hspace{12cm}
\nonumber\\
\hspace{10mm}
\sum_{s'}\int\!d^3r'\left(
-\langle \br,s|\hat h^*|\br',s'\rangle \langle \br',s'|\hat\rho^*|\br'',s''\rangle  
+\langle \br,s|\hat\rho^*|\br',s'\rangle \langle \br',s'|\hat h^*|\br'',s''\rangle \right.
\nonumber\\
\hspace{10mm}
\left.
+\langle \br,s|\hat\Delta^{\dagger}|\br',s'\rangle \langle \br',s'|\hat\kappa|\br'',s''\rangle  
-\langle \br,s|\hat\kappa^{\dagger}|\br',s'\rangle \langle \br',s'|\hat\Delta|\br'',s''\rangle 
\right),
\nonumber\\
i\hbar\langle \br,s|\dot{\hat\kappa}^{\dagger}|\br'',s''\rangle  = \langle \br,s|\hat\Delta^{\dagger}|\br'',s''\rangle 
\hspace{85mm}
\nonumber\\
\hspace{10mm}
+\sum_{s'}\int\!d^3r'\left(
-\langle \br,s|\hat h^*|\br',s'\rangle \langle \br',s'|\hat\kappa^{\dagger}|\br'',s''\rangle  
-\langle \br,s|\hat\kappa^{\dagger}|\br',s'\rangle \langle \br',s'|\hat h|\br'',s''\rangle \right.
\nonumber\\
\hspace{10mm}
\left.
-\langle \br,s|\hat\Delta^{\dagger}|\br',s'\rangle \langle \br',s'|\hat\rho|\br'',s''\rangle  
-\langle \br,s|\hat\rho^*|\br',s'\rangle \langle \br',s'|\hat\Delta^{\dagger}|\br'',s''\rangle 
\right).
\label{HFmatr}
\end{eqnarray}
 We do not specify the isospin indices in order to make
formulae more transparent. They will be re-introduced at the end. 
Let us introduce the more compact notation
$\langle \br,s|\hat X|\br',s'\rangle =X_{rr'}^{ss'}$. Then
the set of TDHFB equations (\ref{HFmatr}) with specified spin indices reads
\begin{eqnarray}
&&i\hbar\dot{\rho}_{rr''}^{\uparrow\uparrow} =
\int\!d^3r'(
 h_{rr'}^{\uparrow\uparrow}\rho_{r'r''}^{\uparrow\uparrow} 
-\rho_{rr'}^{\uparrow\uparrow} h_{r'r''}^{\uparrow\uparrow}
+\hat h_{rr'}^{\uparrow\downarrow}\rho_{r'r''}^{\downarrow\uparrow} 
-\rho_{rr'}^{\uparrow\downarrow} h_{r'r''}^{\downarrow\uparrow}
-\Delta_{rr'}^{\uparrow\downarrow}{\kappa^{\dagger}}_{r'r''}^{\downarrow\uparrow}
+\kappa_{rr'}^{\uparrow\downarrow}{\Delta^{\dagger}}_{r'r''}^{\downarrow\uparrow}),
\nonumber\\
&&i\hbar\dot{\rho}_{rr''}^{\uparrow\downarrow} =
\int\!d^3r'(
 h_{rr'}^{\uparrow\uparrow}\rho_{r'r''}^{\uparrow\downarrow} 
-\rho_{rr'}^{\uparrow\uparrow} h_{r'r''}^{\uparrow\downarrow}
+\hat h_{rr'}^{\uparrow\downarrow}\rho_{r'r''}^{\downarrow\downarrow} 
-\rho_{rr'}^{\uparrow\downarrow} h_{r'r''}^{\downarrow\downarrow}),
\nonumber\\
&&i\hbar\dot{\rho}_{rr''}^{\downarrow\uparrow} =
\int\!d^3r'(
 h_{rr'}^{\downarrow\uparrow}\rho_{r'r''}^{\uparrow\uparrow} 
-\rho_{rr'}^{\downarrow\uparrow} h_{r'r''}^{\uparrow\uparrow}
+\hat h_{rr'}^{\downarrow\downarrow}\rho_{r'r''}^{\downarrow\uparrow} 
-\rho_{rr'}^{\downarrow\downarrow} h_{r'r''}^{\downarrow\uparrow}),
\nonumber\\
&&i\hbar\dot{\rho}_{rr''}^{\downarrow\downarrow} =
\int\!d^3r'(
 h_{rr'}^{\downarrow\uparrow}\rho_{r'r''}^{\uparrow\downarrow} 
-\rho_{rr'}^{\downarrow\uparrow} h_{r'r''}^{\uparrow\downarrow}
+\hat h_{rr'}^{\downarrow\downarrow}\rho_{r'r''}^{\downarrow\downarrow} 
-\rho_{rr'}^{\downarrow\downarrow} h_{r'r''}^{\downarrow\downarrow}
-\Delta_{rr'}^{\downarrow\uparrow}{\kappa^{\dagger}}_{r'r''}^{\uparrow\downarrow}
+\kappa_{rr'}^{\downarrow\uparrow}{\Delta^{\dagger}}_{r'r''}^{\uparrow\downarrow}),
\nonumber\\
&&i\hbar\dot{\kappa}_{rr''}^{\uparrow\downarrow} = -\hat\Delta_{rr''}^{\uparrow\downarrow}
+\int\!d^3r'\left(
 h_{rr'}^{\uparrow\uparrow}\kappa_{r'r''}^{\uparrow\downarrow} 
+\kappa_{rr'}^{\uparrow\downarrow} {h^*}_{r'r''}^{\downarrow\downarrow}
+\Delta_{rr'}^{\uparrow\downarrow}{\rho^*}_{r'r''}^{\downarrow\downarrow} 
+\rho_{rr'}^{\uparrow\uparrow}\Delta_{r'r''}^{\uparrow\downarrow}
\right),
\nonumber\\
&&i\hbar\dot{\kappa}_{rr''}^{\downarrow\uparrow} = -\hat\Delta_{rr''}^{\downarrow\uparrow}
+\int\!d^3r'\left(
 h_{rr'}^{\downarrow\downarrow}\kappa_{r'r''}^{\downarrow\uparrow} 
+\kappa_{rr'}^{\downarrow\uparrow} {h^*}_{r'r''}^{\uparrow\uparrow}
+\Delta_{rr'}^{\downarrow\uparrow}{\rho^*}_{r'r''}^{\uparrow\uparrow} 
+\rho_{rr'}^{\downarrow\downarrow}\Delta_{r'r''}^{\downarrow\uparrow}
\right).
\label{HFsp}
\end{eqnarray}
This set of equations must be complemented by the complex conjugated equations.
Writing these equations, we neglected the diagonal matrix elements in spin,
$\kappa_{rr'}^{ss}$ and $\Delta_{rr'}^{ss}$. It is shown in Appendix A that such 
approximation works very well in the case of monopole pairing considered here.

We will work with the Wigner transform \cite{Ring} of 
equations (\ref{HFsp}). The relevant mathematical details can be found in
\cite{Malov}. The most essential relations are outlined in  Appendix B.
>From now on, we will not write out the coordinate 
dependence $(\br,\bp)$ of all functions in order to make the formulae 
more transparent. The
Wigner transform of (\ref{HFsp}) can be written as
\begin{eqnarray}
      i\hbar\dot f^{\uparrow\uparrow} &=&i\hbar\{h^{\uparrow\uparrow},f^{\uparrow\uparrow}\}
+h^{\uparrow\downarrow}f^{\downarrow\uparrow}-f^{\uparrow\downarrow}h^{\downarrow\uparrow}
+\frac{i\hbar}{2}\{h^{\uparrow\downarrow},f^{\downarrow\uparrow}\}
-\frac{i\hbar}{2}\{f^{\uparrow\downarrow},h^{\downarrow\uparrow}\}
\nonumber\\
&-&\frac{\hbar^2}{8}\{\!\{h^{\uparrow\downarrow},f^{\downarrow\uparrow}\}\!\}
+\frac{\hbar^2}{8}\{\!\{f^{\uparrow\downarrow},h^{\downarrow\uparrow}\}\!\} 
+ \kappa\Delta^* - \Delta\kappa^* 
\nonumber\\
&+&\frac{i\hbar}{2}\{\kappa,\Delta^*\}-\frac{i\hbar}{2}\{\Delta,\kappa^*\}
- \frac{\hbar^2}{8}\{\!\{\kappa,\Delta^*\}\!\} + \frac{\hbar^2}{8}\{\!\{\Delta,\kappa^*\}\!\}
+...,
\nonumber\\
      i\hbar\dot f^{\downarrow\downarrow} &=&i\hbar\{h^{\downarrow\downarrow},f^{\downarrow\downarrow}\}
+h^{\downarrow\uparrow}f^{\uparrow\downarrow}-f^{\downarrow\uparrow}h^{\uparrow\downarrow}
+\frac{i\hbar}{2}\{h^{\downarrow\uparrow},f^{\uparrow\downarrow}\}
-\frac{i\hbar}{2}\{f^{\downarrow\uparrow},h^{\uparrow\downarrow}\}
\nonumber\\
&-&\frac{\hbar^2}{8}\{\!\{h^{\downarrow\uparrow},f^{\uparrow\downarrow}\}\!\}
+\frac{\hbar^2}{8}\{\!\{f^{\downarrow\uparrow},h^{\uparrow\downarrow}\}\!\} 
+ \bar\Delta^* \bar\kappa - \bar\kappa^* \bar\Delta
\nonumber\\
&+&\frac{i\hbar}{2}\{\bar\Delta^*,\bar\kappa\}-\frac{i\hbar}{2}\{\bar\kappa^*,\bar\Delta\}
- \frac{\hbar^2}{8}\{\!\{\bar\Delta^*,\bar\kappa\}\!\} + \frac{\hbar^2}{8}\{\!\{\bar\kappa^*,\bar\Delta\}\!\}
+...,
\nonumber\\
      i\hbar\dot f^{\uparrow\downarrow} &=&
f^{\uparrow\downarrow}(h^{\uparrow\uparrow}-h^{\downarrow\downarrow})
+\frac{i\hbar}{2}\{(h^{\uparrow\uparrow}+h^{\downarrow\downarrow}),f^{\uparrow\downarrow}\}
-\frac{\hbar^2}{8}\{\!\{(h^{\uparrow\uparrow}-h^{\downarrow\downarrow}),f^{\uparrow\downarrow}\}\!\}
\nonumber\\
&-&h^{\uparrow\downarrow}(f^{\uparrow\uparrow}-f^{\downarrow\downarrow})
+\frac{i\hbar}{2}\{h^{\uparrow\downarrow},(f^{\uparrow\uparrow}+f^{\downarrow\downarrow})\}
+\frac{\hbar^2}{8}\{\!\{h^{\uparrow\downarrow},(f^{\uparrow\uparrow}-f^{\downarrow\downarrow})\}\!\}+....,
\nonumber\\
      i\hbar\dot f^{\downarrow\uparrow} &=&
f^{\downarrow\uparrow}(h^{\downarrow\downarrow}-h^{\uparrow\uparrow})
+\frac{i\hbar}{2}\{(h^{\downarrow\downarrow}+h^{\uparrow\uparrow}),f^{\downarrow\uparrow}\}
-\frac{\hbar^2}{8}\{\!\{(h^{\downarrow\downarrow}-h^{\uparrow\uparrow}),f^{\downarrow\uparrow}\}\!\}
\nonumber\\
&-&h^{\downarrow\uparrow}(f^{\downarrow\downarrow}-f^{\uparrow\uparrow})
+\frac{i\hbar}{2}\{h^{\downarrow\uparrow},(f^{\downarrow\downarrow}+f^{\uparrow\uparrow})\}
+\frac{\hbar^2}{8}\{\!\{h^{\downarrow\uparrow},(f^{\downarrow\downarrow}-f^{\uparrow\uparrow})\}\!\}+...,
\nonumber\\
      i\hbar\dot \kappa &=& \kappa\,(h^{\uparrow\uparrow}+\bar h^{\downarrow\downarrow})
  +\frac{i\hbar}{2}\{(h^{\uparrow\uparrow}-\bar h^{\downarrow\downarrow}),\kappa\}    
  -\frac{\hbar^2}{8}\{\!\{(h^{\uparrow\uparrow}+\bar h^{\downarrow\downarrow}),\kappa\}\!\}   
 \nonumber\\ 
 &+&\Delta\,(f^{\uparrow\uparrow}+\bar f^{\downarrow\downarrow})
  +\frac{i\hbar}{2}\{(f^{\uparrow\uparrow}-\bar f^{\downarrow\downarrow}),\Delta\}    
  -\frac{\hbar^2}{8}\{\!\{(f^{\uparrow\uparrow}+\bar f^{\downarrow\downarrow}),\Delta\}\!\}
  - \Delta + ...,
 \nonumber\\     
    i\hbar\dot \kappa^* &=&  -\kappa^*(h^{\uparrow\uparrow}+\bar h^{\downarrow\downarrow}) 
  +\frac{i\hbar}{2}\{(h^{\uparrow\uparrow}-\bar h^{\downarrow\downarrow}),\kappa^*\}    
  +\frac{\hbar^2}{8}\{\!\{(h^{\uparrow\uparrow}+\bar h^{\downarrow\downarrow}),\kappa^*\}\!\}   
  \nonumber\\
  &-& \Delta^*(f^{\uparrow\uparrow}+\bar f^{\downarrow\downarrow}) 
  +\frac{i\hbar}{2}\{(f^{\uparrow\uparrow}-\bar f^{\downarrow\downarrow}),\Delta^*\}    
  +\frac{\hbar^2}{8}\{\!\{(f^{\uparrow\uparrow}+\bar f^{\downarrow\downarrow}),\Delta^*\}\!\}
  + \Delta^* +..., 
\label{WHF}
\end{eqnarray} 
where the functions $h$, $f$, $\Delta$, and $\kappa$ are the Wigner
transforms of $\hat h$, $\hat\rho$, $\hat\Delta$, and $\hat\kappa$,
respectively, $\bar f(\br,\bp)=f(\br,-\bp)$,
 $\{f,g\}$ is the Poisson
bracket of the functions $f(\br,\bp)$ and $g(\br,\bp)$ and
$\{\{f,g\}\}$ is their double Poisson bracket;
the dots stand for terms proportional to higher powers of $\hbar$.
This set of equations must be complemented by the dynamical equations for 
$\bar f^{\uparrow\uparrow}, \bar f^{\downarrow\downarrow}, \bar f^{\uparrow\downarrow}, 
\bar f^{\downarrow\uparrow},\bar\kappa,\bar\kappa^*$.
They are obtained by the change $\bp \rightarrow -\bp$ in arguments of functions and Poisson brackets. 
So, in reality we deal with the set of twelve equations. We introduced the notation
$\kappa \equiv \kappa^{\uparrow\downarrow}$ and $\Delta \equiv \Delta^{\uparrow\downarrow}$.
Symmetry properties of matrices $\hat\kappa, \hat\Delta$ and the properties of their Wigner 
transforms (see Appendix B) allow one to replace the functions 
$\kappa^{\downarrow\uparrow}(\br,\bp)$ and  $\Delta^{\downarrow\uparrow}(\br,\bp)$ by the functions $\bar\kappa^{\uparrow\downarrow}(\br,\bp)$ and  $\bar\Delta^{\uparrow\downarrow}(\br,\bp)$. 

Following the paper \cite{BaMo} we will write above equations in terms of spin-scalar
$$f^+=f^{\uparrow\uparrow}+ f^{\downarrow\downarrow}$$
and spin-vector
$$f^-=f^{\uparrow\uparrow}- f^{\downarrow\downarrow}$$
functions. Furthermore, it is useful to 
rewrite the obtained equations in terms of even and odd functions 
$f_{e}=\frac{1}{2}(f+\bar f)$ and $f_{o}=\frac{1}{2}(f-\bar f)$
and real and imaginary parts of $\kappa$ and $\Delta$: $\kappa^r=\frac{1}{2}(\kappa+\kappa^*),\,
\kappa^i=\frac{1}{2i}(\kappa-\kappa^*),\,\Delta^r=\frac{1}{2}(\Delta+\Delta^*),\,
\Delta^i=\frac{1}{2i}(\Delta-\Delta^*)$. We have

\begin{eqnarray}
i\hbar\dot f^{+}_e &=&\frac{i\hbar}{2}\left[
\{h^+_o,f^+_e\}+\{h^+_e,f^+_o\}+\{h^-_o,f^-_e\}+\{h^-_e,f^-_o\} \right]
\nonumber\\
&+&{i\hbar}\left[
 \{h^{\uparrow\downarrow}_o,f^{\downarrow\uparrow}_e\}
+\{h^{\uparrow\downarrow}_e,f^{\downarrow\uparrow}_o\}
+\{h^{\downarrow\uparrow}_o,f^{\uparrow\downarrow}_e\}
+\{h^{\downarrow\uparrow}_e,f^{\uparrow\downarrow}_o\} \right]
\nonumber\\
&+&4i\left([\kappa^i_e\Delta^r_e]-[\kappa^r_e\Delta^i_e]
+[\kappa^i_o\Delta^r_o]-[\kappa^r_o\Delta^i_o]\right)
+...,
\nonumber\\
i\hbar\dot f^{+}_o &=&\frac{i\hbar}{2}\left[
\{h^+_o,f^+_o\}+\{h^+_e,f^+_e\}+\{h^-_o,f^-_o\}+\{h^-_e,f^-_e\} \right]
\nonumber\\
&+&{i\hbar}\left[
 \{h^{\uparrow\downarrow}_o,f^{\downarrow\uparrow}_o\}
+\{h^{\uparrow\downarrow}_e,f^{\downarrow\uparrow}_e\}
+\{h^{\downarrow\uparrow}_o,f^{\uparrow\downarrow}_o\}
+\{h^{\downarrow\uparrow}_e,f^{\uparrow\downarrow}_e\} \right]
\nonumber\\
&+&2i\hbar\left(
\{\kappa^r_e,\Delta^r_e\}+\{\kappa^i_e,\Delta^i_e\}
+\{\kappa^r_o,\Delta^r_o\}+\{\kappa^i_o,\Delta^i_o\}   \right)
+...,
\nonumber\\
      i\hbar\dot f^{-}_e &=&
 2(h^{\uparrow\downarrow}_e f^{\downarrow\uparrow}_e
+h^{\uparrow\downarrow}_o f^{\downarrow\uparrow}_o
-h^{\downarrow\uparrow}_e f^{\uparrow\downarrow}_e
-h^{\downarrow\uparrow}_o f^{\uparrow\downarrow}_o)
\nonumber\\  
&+&\frac{i\hbar}{2}\left[
\{h^+_o,f^-_e\}+\{h^+_e,f^-_o\}+\{h^-_o,f^+_e\}+\{h^-_e,f^+_o\} \right]
\nonumber\\
&+&2i\hbar\left(
\{\kappa^r_e,\Delta^r_o\}+\{\kappa^i_e,\Delta^i_o\}
+\{\kappa^r_o,\Delta^r_e\}+\{\kappa^i_o,\Delta^i_e\}   \right)
\nonumber\\
&-&\frac{\hbar^2}{4}\left[
 \{\!\{h^{\uparrow\downarrow}_e,f^{\downarrow\uparrow}_e\}\!\}
+\{\!\{h^{\uparrow\downarrow}_o,f^{\downarrow\uparrow}_o\}\!\} 
-\{\!\{h^{\downarrow\uparrow}_e,f^{\uparrow\downarrow}_e\}\!\}
-\{\!\{h^{\downarrow\uparrow}_o,f^{\uparrow\downarrow}_o\}\!\} \right]
+...,
\nonumber\\
      i\hbar\dot f^{-}_o &=&
 2(h^{\uparrow\downarrow}_e f^{\downarrow\uparrow}_o
+h^{\uparrow\downarrow}_o f^{\downarrow\uparrow}_e
-h^{\downarrow\uparrow}_e f^{\uparrow\downarrow}_o
-h^{\downarrow\uparrow}_o f^{\uparrow\downarrow}_e)
\nonumber\\
&+&4i\left(
[\kappa^i_e\Delta^r_o]-[\kappa^r_e\Delta^i_o]
+[\kappa^i_o\Delta^r_e]-[\kappa^r_o\Delta^i_e]\right)
\nonumber\\  
&+&\frac{i\hbar}{2}\left[
\{h^+_o,f^-_o\}+\{h^+_e,f^-_e\}+\{h^-_o,f^+_o\}+\{h^-_e,f^+_e\} \right]
\nonumber\\
&-&\frac{\hbar^2}{4}\left[
 \{\!\{h^{\uparrow\downarrow}_e,f^{\downarrow\uparrow}_o\}\!\}
+\{\!\{h^{\uparrow\downarrow}_o,f^{\downarrow\uparrow}_e\}\!\} 
-\{\!\{h^{\downarrow\uparrow}_e,f^{\uparrow\downarrow}_o\}\!\}
-\{\!\{h^{\downarrow\uparrow}_o,f^{\uparrow\downarrow}_e\}\!\} \right]
+...,
\nonumber\\
      i\hbar\dot f^{\uparrow\downarrow}_e &=&
 \left[ h^-_e f^{\uparrow\downarrow}_e + h^-_o f^{\uparrow\downarrow}_o 
- h^{\uparrow\downarrow}_e f^-_e
- h^{\uparrow\downarrow}_o f^-_o  \right]
\nonumber\\
 &+&\frac{i\hbar}{2}\left[
 \{h^{\uparrow\downarrow}_e,f^+_o\}
+\{h^{\uparrow\downarrow}_o,f^+_e\}
+\{h^+_e,f^{\uparrow\downarrow}_o\} + \{h^+_o,f^{\uparrow\downarrow}_e\} 
 \right]
\nonumber\\
&+&\frac{\hbar^2}{8}\left[
\{\!\{h^{\uparrow\downarrow}_e,f^-_e\}\!\} +\{\!\{h^{\uparrow\downarrow}_o,f^-_o\}\!\}
-\{\!\{h^-_e,f^{\uparrow\downarrow}_e\}\!\} - \{\!\{h^-_o,f^{\uparrow\downarrow}_o\}\!\} 
 \right]
+...,
\nonumber\\  
      i\hbar \dot f^{\downarrow\uparrow}_e &=&       
 -\left[
h^-_e f^{\downarrow\uparrow}_e 
+h^-_o f^{\downarrow\uparrow}_o 
-h^{\downarrow\uparrow}_e f^-_e
-h^{\downarrow\uparrow}_o f^-_o  \right]
\nonumber\\
 &+&\frac{i\hbar}{2}\left[
 \{h^{\downarrow\uparrow}_e,f^+_o\}
+\{h^{\downarrow\uparrow}_o,f^+_e\}
+\{h^+_e,f^{\downarrow\uparrow}_o\} 
+\{h^+_o,f^{\downarrow\uparrow}_e\} 
 \right]
\nonumber\\
&-&\frac{\hbar^2}{8}\left[
 \{\!\{h^{\downarrow\uparrow}_e,f^-_e\}\!\} 
+\{\!\{h^{\downarrow\uparrow}_o,f^-_o\}\!\}
-\{\!\{h^-_e,f^{\downarrow\uparrow}_e\}\!\} 
-\{\!\{h^-_o,f^{\downarrow\uparrow}_o\}\!\} 
 \right]
+...,
\nonumber\\
      i\hbar\dot f^{\uparrow\downarrow}_o &=&
 \left[ h^-_e f^{\uparrow\downarrow}_o + h^-_o f^{\uparrow\downarrow}_e 
- h^{\uparrow\downarrow}_e f^-_o
- h^{\uparrow\downarrow}_o f^-_e  \right]
\nonumber\\
 &+&\frac{i\hbar}{2}\left[
 \{h^{\uparrow\downarrow}_e,f^+_e\}
+\{h^{\uparrow\downarrow}_o,f^+_o\}
+\{h^+_e,f^{\uparrow\downarrow}_e\} + \{h^+_o,f^{\uparrow\downarrow}_o\} 
 \right]
\nonumber\\
&+&\frac{\hbar^2}{8}\left[
\{\!\{h^{\uparrow\downarrow}_e,f^-_o\}\!\} +\{\!\{h^{\uparrow\downarrow}_o,f^-_e\}\!\}
-\{\!\{h^-_e,f^{\uparrow\downarrow}_o\}\!\} - \{\!\{h^-_o,f^{\uparrow\downarrow}_e\}\!\} 
 \right]
+...,
\nonumber\\  
      i\hbar \dot f^{\downarrow\uparrow}_o &=&       
 -\left[
h^-_e f^{\downarrow\uparrow}_o
+h^-_o f^{\downarrow\uparrow}_e 
-h^{\downarrow\uparrow}_e f^-_o
-h^{\downarrow\uparrow}_o f^-_e  \right]
\nonumber\\
 &+&\frac{i\hbar}{2}\left[
 \{h^{\downarrow\uparrow}_e,f^+_e\}
+\{h^{\downarrow\uparrow}_o,f^+_o\}
+\{h^+_e,f^{\downarrow\uparrow}_e\} 
+\{h^+_o,f^{\downarrow\uparrow}_o\} 
 \right]
\nonumber\\
&-&\frac{\hbar^2}{8}\left[
 \{\!\{h^{\downarrow\uparrow}_e,f^-_o\}\!\} 
+\{\!\{h^{\downarrow\uparrow}_o,f^-_e\}\!\}
-\{\!\{h^-_e,f^{\downarrow\uparrow}_o\}\!\} 
-\{\!\{h^-_o,f^{\downarrow\uparrow}_e\}\!\} 
 \right]
+...,
\nonumber\\
     i\hbar\dot \kappa^r_e &=& i[h^+_e\,\kappa^i_e+h^-_o\,\kappa^i_o]
  +\frac{i\hbar}{2}\{h^+_o,\kappa^r_e+h^-_e,\kappa^r_o\}      
 \nonumber\\ 
 &+&i[f^+_e\,\Delta^i_e+ f^-_o\,\Delta^i_o]
  +\frac{i\hbar}{2}\{f^+_o,\Delta^r_e+ f^-_e,\Delta^r_o\} -i\Delta^i_e
  +...,
\nonumber\\
     i\hbar\dot \kappa^r_o &=& i[h^+_e\,\kappa^i_o+h^-_o\,\kappa^i_e]
  +\frac{i\hbar}{2}\{h^+_o,\kappa^r_o+h^-_e,\kappa^r_e\}      
 \nonumber\\ 
 &+&i[f^+_e\,\Delta^i_o+ f^-_o\,\Delta^i_e]
  +\frac{i\hbar}{2}\{f^+_o,\Delta^r_o+ f^-_e,\Delta^r_e\} -i\Delta^i_o
  +...,
\nonumber\\
     i\hbar\dot \kappa^i_e &=& -i[h^+_e\,\kappa^r_e+h^-_o\,\kappa^r_o]
  +\frac{i\hbar}{2}\{h^+_o,\kappa^i_e+h^-_e,\kappa^i_o\}      
 \nonumber\\ 
 &-&i[f^+_e\,\Delta^r_e+ f^-_o\,\Delta^r_o]
  +\frac{i\hbar}{2}\{f^+_o,\Delta^i_e+ f^-_e,\Delta^i_o\} +i\Delta^r_e
  +...,
\nonumber\\
     i\hbar\dot \kappa^i_o &=& -i[h^+_e\,\kappa^r_o+h^-_o\,\kappa^r_e]
  +\frac{i\hbar}{2}\{h^+_o,\kappa^i_o+h^-_e,\kappa^i_e\}      
 \nonumber\\ 
 &-&i[f^+_e\,\Delta^r_o+ f^-_o\,\Delta^r_e]
  +\frac{i\hbar}{2}\{f^+_o,\Delta^i_o+ f^-_e,\Delta^i_e\} +i\Delta^r_o
  +...,
 \label{WHFeo}
\end{eqnarray}
The following notation is introduced here:
 $h^{\pm}=h^{\uparrow\uparrow}\pm h^{\downarrow\downarrow},\quad
  [ab]=ab-\frac{\hbar^2}{8}\{\!\{a,b\}\!\},\quad [ab+cd+...]=[ab]+[cd]+...,\quad
\{a,b+c,d+...\}=\{a,b\}+\{c,d\}+...  $.

These twelve equations will be solved by the method of moments in a small amplitude 
approximation. To this end 
all functions $f(\br,\bp,t)$ and $\kappa(\br,\bp,t)$ are divided into equilibrium part 
and deviation (variation): $f(\br,\bp,t)=f(\br,\bp)_{eq}+\delta f(\br,\bp,t)$, 
$\kappa(\br,\bp,t)=\kappa(\br,\bp)_{eq}+\delta \kappa(\br,\bp,t)$.
Then equations are linearized  neglecting quadratic terms.

From general arguments one can expect that the phase of $\Delta$ (and
of $\kappa$, since both are linked, according to equation (\ref{DK}))
is much more relevant than its magnitude, since the former determines
the superfluid velocity. After linearization, the phase of $\Delta$
(and of $\kappa$) is expressed by $\delta\Delta^i$ (and $\delta\kappa^i$), while
$\delta\Delta^r$ (and $\delta\kappa^r$) describes oscillations of the
magnitude of $\Delta$ (and of $\kappa$). Let us therefore assume that
\begin{equation}
\delta\kappa^r(\br,\bp)\ll\delta\kappa^i(\br,\bp).
\label{approx1}
\end{equation}
This assumption was explicitly confirmed in \cite{M.Urban} for the case
of superfluid trapped fermionic atoms, where it was shown that
$\delta\Delta^r$ is suppressed with respect to $\delta\Delta^i$ by one
order of $\Delta/E_{\rm F}$, where $E_{\rm F}$ denotes the Fermi energy.

The assumption (\ref{approx1}) allows one to neglect all terms containing the variations
$\delta \kappa^r$ and $\delta \Delta^r$ in the equations 
(\ref{WHFeo}) after their linearization. In this case the "small" variations 
$\delta \kappa^r$ and $\delta \Delta^r$  
will not affect the dynamics of the "big" variations 
$\delta \kappa^i$ and $\delta \Delta^i$ . This means that the
dynamical equations for the "big" variations can be considered
independently from that of the "small" variations, and we will finally
deal with a set of only ten equations.

\section{Model Hamiltonian}

 The microscopic Hamiltonian of the model, harmonic oscillator with 
spin orbit potential plus separable quadrupole-quadrupole and 
spin-spin residual interactions is given by
\begin{eqnarray}
\label{Ham}
 H=\sum\limits_{i=1}^A\left[\frac{\hat\bp_i^2}{2m}+\frac{1}{2}m\omega^2\br_i^2
-\eta\hat \bl_i\hat \bS_i\right]+H_{qq}+H_{ss}
\end{eqnarray}
with
\begin{eqnarray}
\label{Hqq}
&& H_{qq}=\!
\sum_{\mu=-2}^{2}(-1)^{\mu}
\left\{\bar{\kappa}
 \sum\limits_i^Z\!\sum\limits_j^N
+\frac{\kappa}{2}
\left[\sum\limits_{i,j(i\neq j)}^{Z}
+\sum\limits_{i,j(i\neq j)}^{N}
\right]
\right\}
q_{2-\mu}(\br_i)q_{2\mu}(\br_j)
,
\\
\label{Hss}
&&H_{ss}=\!
\sum_{\mu=-1}^{1}(-1)^{\mu}
\left\{\bar{\chi}
 \sum\limits_i^Z\!\sum\limits_j^N
+\frac{\chi}{2}
\left[
\sum\limits_{i,j(i\neq j)}^{Z}
+\sum\limits_{i,j(i\neq j)}^{N}
\right]
\right\}
\hat S_{-\mu}(i)\hat S_{\mu}(j)
\,\delta(\br_i-\br_j),
\end{eqnarray}
where $N$ and $Z$ are the numbers of neutrons and protons
and $\hat S_{\mu}$ are spin matrices \cite{Var}:
\begin{equation}
\hat S_1=-\frac{\hbar}{\sqrt2}{0\quad 1\choose 0\quad 0},\quad
\hat S_0=\frac{\hbar}{2}{1\quad\, 0\choose 0\, -\!1},\quad
\hat S_{-1}=\frac{\hbar}{\sqrt2}{0\quad 0\choose 1\quad 0}.
\label{S}
\end{equation}

\subsection{Mean Field}

Let us analyze the mean field generated by this Hamiltonian.

\subsubsection{Spin-orbit Potential}

Written in cyclic coordinates, the spin orbit part of the
Hamiltonian reads
$$\hat h_{ls}=-\eta\sum_{\mu=-1}^1(-)^{\mu}\hat l_{\mu}\hat S_{-\mu}
=-\eta{\quad\hat l_0\frac{\hbar}{2}\quad\; \hat l_{-1}\frac{\hbar}{\sqrt2} \choose 
 -\hat l_{1}\frac{\hbar}{\sqrt2}\; -\hat l_0\frac{\hbar}{2}},
$$
where \cite{Var}
\begin{equation}
\label{lqu}
\hat l_{\mu}=-\hbar\sqrt2\sum_{\nu,\alpha}C_{1\nu,1\alpha}^{1\mu}r_{\nu}\nabla_{\alpha},
\end{equation}
cyclic coordinates $r_{-1}, r_0, r_1$ 
are defined in \cite{Var},
$C_{1\sigma,1\nu}^{\lambda\mu}$ is a Clebsch-Gordan
coefficient and
\begin{eqnarray}
&&\hat l_1=\hbar(r_0\nabla_1-r_1\nabla_0)=
-\frac{1}{\sqrt2}(\hat l_x+i\hat l_y),\quad
\hat l_0=\hbar(r_{-1}\nabla_1-r_1\nabla_{-1})=\hat l_z,
\nonumber\\
&&\hat l_{-1}=\hbar(r_{-1}\nabla_0-r_0\nabla_{-1})=
\frac{1}{\sqrt2}(\hat l_x-i\hat l_y),
\nonumber\\
&&\hat l_x=-i\hbar(y\nabla_z-z\nabla_y),\quad
\hat l_y=-i\hbar(z\nabla_x-x\nabla_z),\quad
\hat l_z=-i\hbar(x\nabla_y-y\nabla_x).
\label{lxyz}
\end{eqnarray}
 Matrix elements of $\hat h_{ls}$ in coordinate space can be obviously written \cite{BaMo} as
\begin{eqnarray}
\langle \br_1,s_1|\hat h_{ls}|\br_2,s_2\rangle 
&=&-\frac{\hbar}{2}\eta\left[\hat l_{0}(\br_1)(\delta_{s_1\uparrow}\delta_{s_2\uparrow}
-\delta_{s_1\downarrow}\delta_{s_2\downarrow})\right.
\nonumber\\
&&+\left.\sqrt2\, \hat l_{-1}(\br_1)\delta_{s_1\uparrow}\delta_{s_2\downarrow}
-\sqrt2\, \hat l_{1}(\br_1)\delta_{s_1\downarrow}\delta_{s_2\uparrow}\right]\delta(\br_1-\br_2).
\label{Hrr'}
\end{eqnarray}
Their Wigner transform reads \cite{BaMo}:
\begin{eqnarray}
 h_{ls}^{s_1s_2}(\br,\bp)
&=&-\frac{\hbar}{2}\eta\left[l_{0}(\br,\bp)(\delta_{s_1\uparrow}\delta_{s_2\uparrow}
-\delta_{s_1\downarrow}\delta_{s_2\downarrow})\right.
\nonumber\\
&&+\left.\sqrt2 l_{-1}(\br,\bp)\delta_{s_1\uparrow}\delta_{s_2\downarrow}
-\sqrt2 l_{1}(\br,\bp)\delta_{s_1\downarrow}\delta_{s_2\uparrow}\right],
\label{Hrp}
\end{eqnarray}
where
$l_{\mu}=-i\sqrt2\sum_{\nu,\alpha}C_{1\nu,1\alpha}^{1\mu}r_{\nu}p_{\alpha}$.

\subsubsection{Quadrupole-quadrupole interaction}

 The contribution of $H_{qq}$ to the mean field potential is easily
found by replacing one of the $q_{2\mu}$ operators by the average value.
We have
\begin{equation}
\label{potenirr}
V^{\tau}_{qq}=\sqrt6\sum_{\mu}(-1)^{\mu}Z_{2-\mu}^{\tau +}q_{2\mu}.
\end{equation}
 Here
\begin{equation}
\label{Z2mu}
Z_{2\mu}^{n+}=\kappa R_{2\mu}^{n+}
+\bar{\kappa}R_{2\mu}^{p+}\,,\quad
Z_{2\mu}^{p+}=\kappa R_{2\mu}^{p+}
+\bar{\kappa}R_{2\mu}^{n+},\quad  R_{2\mu}^{\tau+}(t)=
\frac{1}{\sqrt6}\int d(\bp,\br)
q_{2\mu}(\br)f^{\tau+}(\br,\bp,t)
\end{equation}
 with
 $\int\! d(\bp,\br)\equiv
(2\pi\hbar)^{-3}\int\! d^3p\,\int\! d^3r$ and $\tau$ being the isospin index.

\subsubsection{Spin-spin interaction}

The analogous expression for $H_{ss}$ is found in a
standard way \cite{BaMoPRC} with the following result for
the Wigner transform of the proton mean field:
\begin{eqnarray}
\label{Vp}
V_{p}^{s s'}(\br,t)&=&
3\chi\frac{\hbar^2}{8}
\left[
\delta_{s\downarrow}\delta_{s'\uparrow}n_p^{\downarrow\uparrow}+
\delta_{s\uparrow}\delta_{s'\downarrow}n_p^{\uparrow\downarrow}
-\delta_{s\downarrow}\delta_{s'\downarrow}n_p^{\uparrow\uparrow}
-\delta_{s\uparrow}\delta_{s'\uparrow}n_p^{\downarrow\downarrow}
\right]
\nonumber\\
&+&\bar\chi\frac{\hbar^2}{8}
\left[
2\delta_{s\downarrow}\delta_{s'\uparrow}n_n^{\downarrow\uparrow}+
2\delta_{s\uparrow}\delta_{s'\downarrow}n_n^{\uparrow\downarrow}
+(\delta_{s\uparrow}\delta_{s'\uparrow}-
\delta_{s\downarrow}\delta_{s'\downarrow})(n_n^{\uparrow\uparrow}-
n_n^{\downarrow\downarrow})
\right],
\end{eqnarray}
where 
${\di n_{\tau}^{ss'}(\br,t)=\int\frac{d^3p}{(2\pi\hbar)^3}f^{ss'}_{\tau}(\br,\bp,t)}$.
The Wigner transform of the neutron mean field $V_n^{ss'}$ is 
obtained from (\ref{Vp}) by the obvious change of indices $p\leftrightarrow n$.

\subsection{Pair potential}

The Wigner transform of the pair potential (pairing gap) $\Delta(\br,\bp)$ is related to 
the Wigner transform of the anomalous density by \cite{Ring}
\begin{equation}
\Delta(\br,\bp)=-\int\! \frac{d^3p'}{(2\pi\hbar)^3}
v(|\bp-\bp'|)\kappa(\br,\bp'),
\label{DK}
\end{equation}
where $v(p)$ is a Fourier transform of the two-body interaction.
We take for the pairing interaction a simple Gaussian of strength $V_0$ 
and range $r_p$  \cite{Ring}
\begin{equation}
v(p)=\beta e^{-\alpha p^2}\!,
\label{v_p}
\end{equation}
with $\beta=-|V_0|(r_p\sqrt{\pi})^3$ and $\alpha=r_p^2/4\hbar^2$. For the values of the parameters, see section 5.1.

\section{Equations of motion}

 Integrating the set of equations (\ref{WHFeo}) over phase space 
with the weights 
\begin{equation}
W =\{r\otimes p\}_{\lambda\mu},\,\{r\otimes r\}_{\lambda\mu},\,
\{p\otimes p\}_{\lambda\mu}, \mbox{ and } 1
\label{weightfunctions}
\end{equation}
one gets dynamic equations for 
the following collective variables:
\begin{eqnarray}
&&\L^{\tau\varsigma}_{\lambda\mu}(t)=\int\! d(\bp,\br) \{r\otimes p\}_{\lambda\mu}
\delta f^{\tau\varsigma}_o(\br,\bp,t),\quad
\R^{\tau\varsigma}_{\lambda\mu}(t)=\int\! d(\bp,\br) \{r\otimes r\}_{\lambda\mu}
\delta f^{\tau\varsigma}_e(\br,\bp,t),\quad
\nonumber\\
&&\P^{\tau\varsigma}_{\lambda\mu}(t)=\int\! d(\bp,\br) \{p\otimes p\}_{\lambda\mu}
\delta f^{\tau\varsigma}_e(\br,\bp,t),\quad
\F^{\tau\varsigma}(t)=\int\! d(\bp,\br)
\delta f^{\tau\varsigma}_e(\br,\bp,t),\quad
\nonumber\\
&&\tilde{\L}^{\tau}_{\lambda\mu}(t)=\int\! d(\bp,\br) \{r\otimes p\}_{\lambda\mu}
\delta \kappa^{\tau i}_o(\br,\bp,t),\quad
\tilde{\R}^{\tau}_{\lambda\mu}(t)=\int\! d(\bp,\br) \{r\otimes r\}_{\lambda\mu}
\delta \kappa^{\tau i}_e(\br,\bp,t),\quad
\nonumber\\
&&\tilde{\P}^{\tau}_{\lambda\mu}(t)=\int\! d(\bp,\br) \{p\otimes p\}_{\lambda\mu}
\delta \kappa^{\tau i}_e(\br,\bp,t),\quad
\label{Varis}
\end{eqnarray}
 where 
$\varsigma\!=+,\,-,\,\uparrow\downarrow,\,\downarrow\uparrow,$
and $\displaystyle \quad
\{r\otimes r\}_{\lambda\mu}=\sum\limits_{\sigma,\nu}
C_{1\sigma,1\nu}^{\lambda\mu}r_{\sigma}r_{\nu}.$

The required expressions for 
$h^{\pm}$, $h^{\uparrow\downarrow}$ and $h^{\downarrow\uparrow}$ are
$$h_{\tau}^{+}=\frac{p^2}{m}+m\,\omega^2r^2
+12\sum_{\mu}(-1)^{\mu}Z_{2\mu}^{\tau+}(t)\{r\otimes r\}_{2-\mu}
+V_{\tau}^+(\br,t)-\mu^{\tau},$$
$\mu^{\tau}$ being the chemical potential of protons ($\tau=p$) or neutrons ($\tau=n$),
$$h_{\tau}^-=-\hbar\eta l_0+V_{\tau}^-(\br,t),\quad
h_{\tau}^{\uparrow\downarrow}=-\frac{\hbar}{\sqrt2}\eta l_{-1}+V_{\tau}^{\uparrow\downarrow}(\br,t),
\quad h_{\tau}^{\downarrow\uparrow}=\frac{\hbar}{\sqrt2}\eta l_{1}+V_{\tau}^{\downarrow\uparrow}(\br,t),
$$
where according to (\ref{Vp})
\begin{eqnarray}
\label{Vss}
V_p^+(\br,t)=-3\frac{\hbar^2}{8}\chi n_p^+(\br,t),\quad
V_p^-(\br,t)=3\frac{\hbar^2}{8}\chi n_p^-(\br,t)+\frac{\hbar^2}{4}\bar\chi n_n^-(\br,t),
\nonumber\\
V_p^{\uparrow\downarrow}(\br,t)=3\frac{\hbar^2}{8}\chi n_p^{\uparrow\downarrow}(\br,t)
+\frac{\hbar^2}{4}\bar\chi n_n^{\uparrow\downarrow}(\br,t),\quad
V_p^{\downarrow\uparrow}(\br,t)=3\frac{\hbar^2}{8}\chi n_p^{\downarrow\uparrow}(\br,t)
+\frac{\hbar^2}{4}\bar\chi n_n^{\downarrow\uparrow}(\br,t)
\end{eqnarray}
and the neutron potentials $V_n^{\varsigma}$ are
obtained by the obvious change of indices $p\leftrightarrow n$.
Variations of these mean fields read:
$$\delta h_{\tau}^{+}=12\sum_{\mu}(-1)^{\mu}\delta Z_{2\mu}^{\tau+}(t)\{r\otimes r\}_{2-\mu}
+\delta V_{\tau}^+(\br,t),$$
where
$\quad \delta Z_{2\mu}^{p+}=\kappa \delta R_{2\mu}^{p+}
+\bar{\kappa}\delta R_{2\mu}^{n+},\quad  \delta R_{\lambda\mu}^{\tau+}(t)\equiv
\R_{\lambda\mu}^{\tau+}(t) \quad$ and
$$
\delta V_p^+(\br,t)=-3\frac{\hbar^2}{8}\chi \delta n_p^+(\br,t),\quad
{\di \delta n_{p}^{+}(\br,t)=\int\frac{d^3p}{(2\pi\hbar)^3}\delta f^{+}_{p}(\br,\bp,t)}.$$
Variations of
$h^{-}$, $h^{\uparrow\downarrow}$ and $h^{\downarrow\uparrow}$ are obtained in a similar way.
Variation of the pair potential is
\begin{equation}
\delta \Delta(\br,\bp,t)=-\int\! \frac{d^3p'}{(2\pi\hbar)^3}
v(|\bp-\bp'|)\delta \kappa(\br,\bp',t).
\label{DKvar}
\end{equation}

We are interested in the scissors mode with quantum number
$K^{\pi}=1^+$. Therefore, we only need the part of dynamic equations 
with $\mu=1$. 

  It is convenient to rewrite the dynamical equations in terms
of isoscalar and isovector variables
\begin{eqnarray}
\label{Isovs}
\bar \R_{\lambda\mu}=\R_{\lambda\mu}^{n}+\R_{\lambda\mu}^{p}
,\quad
\bar \P_{\lambda\mu}=\P_{\lambda\mu}^{n}+\P_{\lambda\mu}^{p}
,\quad
\bar \L_{\lambda\mu}=\L_{\lambda\mu}^{n}+\L_{\lambda\mu}^{p}.
\nonumber\\
\R_{\lambda\mu}=\R_{\lambda\mu}^{n}-\R_{\lambda\mu}^{p},\quad
\P_{\lambda\mu}=\P_{\lambda\mu}^{n}-\P_{\lambda\mu}^{p},\quad
\L_{\lambda\mu}=\L_{\lambda\mu}^{n}-\L_{\lambda\mu}^{p},
\end{eqnarray}
It also is natural to define isovector and isoscalar strength constants
$\kappa_1=\frac{1}{2}(\kappa-\bar\kappa)$ and
$\kappa_0=\frac{1}{2}(\kappa+\bar\kappa)$ connected by the relation
$\kappa_1=\alpha\kappa_0$ \cite{BaSc}.
Then the equations for the neutron and proton systems are transformed
into isovector and isoscalar ones. Supposing that all equilibrium
characteristics of the proton system are equal to that of the neutron
system one decouples isovector and isoscalar equations. This 
approximations looks rather crude, nevertheless the possible 
corrections to it are very small, being of the order 
$(\frac{N-Z}{A})^2$.
The integration yields the following set of equations for {\bf isovector} variables:
\begin{eqnarray}
\label{iv}
     \dot {\L}^{+}_{21}&=&
\frac{1}{m}\P_{21}^{+}-
\left[m\,\omega^2
-4\sqrt3\alpha\kappa_0R_{00}^{\rm eq}
+\sqrt6(1+\alpha)\kappa_0 R_{20}^{\rm eq}\right]\R^{+}_{21}
-i\hbar\frac{\eta}{2}\left[\L_{21}^-
+2\L^{\uparrow\downarrow}_{22}+
\sqrt6\L^{\downarrow\uparrow}_{20}\right],
\nonumber\\
     \dot {\L}^{-}_{21}&=&
\frac{1}{m}\P_{21}^{-}
-
\left[m\,\omega^2+\sqrt6\kappa_0 R_{20}^{\rm eq}
-\frac{\sqrt{3}}{20}\hbar^2 
\left( \chi-\frac{\bar\chi}{3} \right)
\left(\frac{I_1}{a_0^2}+\frac{I_1}{a_1^2}\right)\left(\frac{a_1^2}{{\cal A}_2}-\frac{a_0^2}{{\cal A}_1}\right)
\right]\R^{-}_{21}
-i\hbar\frac{\eta}{2}\L_{21}^+ 
\nonumber\\
&&+\frac{4}{\hbar}|V_0| I_{rp}^{\kappa\Delta}(r') {\tilde\L}_{21},
\nonumber\\
     \dot {\L}^{\uparrow\downarrow}_{22}&=&
\frac{1}{m}\P_{22}^{\uparrow\downarrow}-
\left[m\,\omega^2-2\sqrt6\kappa_0R_{20}^{\rm eq}
-\frac{\sqrt{3}}{5}\hbar^2 
\left( \chi-\frac{\bar\chi}{3} \right)\frac{I_1}{{\cal A}_2}
\right]\R^{\uparrow\downarrow}_{22}
-i\hbar\frac{\eta}{2}\L_{21}^+,
\nonumber\\
     \dot {\L}^{\downarrow\uparrow}_{20}&=&
\frac{1}{m}\P_{20}^{\downarrow\uparrow}-
\left[m\,\omega^2
+2\sqrt6\kappa_0 R_{20}^{\rm eq}\right]\R^{\downarrow\uparrow}_{20}
+\frac{2}{\sqrt3}\kappa_0 R_{20}^{\rm eq}\,\R^{\downarrow\uparrow}_{00}
-i\hbar\frac{\eta}{2}\sqrt{\frac{3}{2}}\L_{21}^+ 
\nonumber\\
&&+\frac{\sqrt{3}}{15}\hbar^2 
\left( \chi-\frac{\bar\chi}{3} \right)I_1 \,
\left[
\left(\frac{1}{{\cal A}_2}-\frac{2}{{\cal A}_1}\right)
\R_{20}^{\downarrow\uparrow}+
\sqrt2
\left(\frac{1}{{\cal A}_2}+\frac{1}{{\cal A}_1}\right)
\R_{00}^{\downarrow\uparrow}
\right],
\nonumber\\
     \dot {\L}^{+}_{11}&=&
-3\sqrt6(1-\alpha)\kappa_0 R_{20}^{\rm eq}\,\R^{+}_{21}
-i\hbar\frac{\eta}{2}\left[\L_{11}^- 
+\sqrt2\L^{\downarrow\uparrow}_{10}\right],
\nonumber\\
     \dot {\L}^{-}_{11}&=&
-\left[3\sqrt6\kappa_0 R_{20}^{\rm eq}
-\frac{\sqrt{3}}{20}\hbar^2 
\left( \chi-\frac{\bar\chi}{3} \right)
\left(\frac{I_1}{a_0^2}-\frac{I_1}{a_1^2}\right)\left(\frac{a_1^2}{{\cal A}_2}-\frac{a_0^2}{{\cal A}_1}\right)
\right]\R^{-}_{21}
-\hbar\frac{\eta}{2}\left[i\L_{11}^+
+\hbar F^{\downarrow\uparrow}\right]
\nonumber\\
&&+\frac{4}{\hbar}|V_0| I_{rp}^{\kappa\Delta}(r') {\tilde\L}_{11},
\nonumber\\
     \dot {\L}^{\downarrow\uparrow}_{10}&=&
-\hbar\frac{\eta}{2\sqrt2}\left[i\L_{11}^+
+\hbar F^{\downarrow\uparrow}\right],
\nonumber\\
     \dot { F}^{\downarrow\uparrow}&=&
-\eta\left[\L_{11}^- +\sqrt2\L^{\downarrow\uparrow}_{10}\right],
\nonumber\\
     \dot {\R}^{+}_{21}&=&
\frac{2}{m}\L_{21}^{+}
-i\hbar\frac{\eta}{2}\left[\R_{21}^-
+2\R^{\uparrow\downarrow}_{22}+
\sqrt6\R^{\downarrow\uparrow}_{20}\right],
\nonumber\\
     \dot {\R}^{-}_{21}&=&
\frac{2}{m}\L_{21}^{-}
-i\hbar\frac{\eta}{2}\R_{21}^+,
\nonumber\\
     \dot {\R}^{\uparrow\downarrow}_{22}&=&
\frac{2}{m}\L_{22}^{\uparrow\downarrow}
-i\hbar\frac{\eta}{2}\R_{21}^+,
\nonumber\\
     \dot {\R}^{\downarrow\uparrow}_{20}&=&
\frac{2}{m}\L_{20}^{\downarrow\uparrow}
-i\hbar\frac{\eta}{2}\sqrt{\frac{3}{2}}\R_{21}^+,
\nonumber\\
     \dot {\P}^{+}_{21}&=&
-2\left[m\,\omega^2+\sqrt6\kappa_0 R_{20}^{\rm eq}\right]\L^{+}_{21}
+6\sqrt6\kappa_0 R_{20}^{\rm eq}\L^{+}_{11}
-i\hbar\frac{\eta}{2}\left[\P_{21}^- 
+2\P^{\uparrow\downarrow}_{22}+\sqrt6\P^{\downarrow\uparrow}_{20}\right]
\nonumber\\
&&+\frac{3\sqrt{3}}{4}\hbar^2 
\chi \frac{I_2}{{\cal A}_1{\cal A}_2}
\left[\left({\cal A}_1-{\cal A}_2\right) \L_{21}^{+} +
\left({\cal A}_1+{\cal A}_2\right) \L_{11}^{+}\right]
\nonumber\\
&&+\frac{4}{\hbar}|V_0| I_{pp}^{\kappa\Delta}(r') {\tilde\P}_{21},
\nonumber\\
     \dot {\P}^{-}_{21}&=&
-2\left[m\,\omega^2+\sqrt6\kappa_0 R_{20}^{\rm eq}\right]\L^{-}_{21}
+6\sqrt6\kappa_0 R_{20}^{\rm eq}\L^{-}_{11}
-6\sqrt2\kappa_0 L_{10}^-(\rm eq)\R^{+}_{21}
-i\hbar\frac{\eta}{2}\P_{21}^{+}
\nonumber\\
&&+\frac{3\sqrt{3}}{4}\hbar^2 
\chi \frac{I_2}{{\cal A}_1{\cal A}_2}
\left[\left({\cal A}_1-{\cal A}_2\right)\L_{21}^{-} +
\left({\cal A}_1+{\cal A}_2\right) \L_{11}^{-}\right],
\nonumber\\
     \dot {\P}^{\uparrow\downarrow}_{22}&=&
-\left[2m\,\omega^2-4\sqrt6\kappa_0 R_{20}^{\rm eq}
-\frac{3\sqrt{3}}{2}\hbar^2 
\chi \frac{I_2}{{\cal A}_2}
\right]\L^{\uparrow\downarrow}_{22}
-i\hbar\frac{\eta}{2}\P_{21}^{+}
,
\nonumber\\
     \dot {\P}^{\downarrow\uparrow}_{20}&=&
-\left[2m\,\omega^2+4\sqrt6\kappa_0 R_{20}^{\rm eq}\right]\L^{\downarrow\uparrow}_{20}
+8\sqrt3\kappa_0 R_{20}^{\rm eq}\L^{\downarrow\uparrow}_{00}
-i\hbar\frac{\eta}{2}\sqrt{\frac{3}{2}}\P_{21}^{+}
\nonumber\\
&&+\frac{\sqrt{3}}{2}\hbar^2 
\chi \frac{I_2}{{\cal A}_1{\cal A}_2}
\left[\left({\cal A}_1-2{\cal A}_2\right)\L_{20}^{\downarrow\uparrow}+
\sqrt2\left({\cal A}_1+{\cal A}_2\right) \L_{00}^{\downarrow\uparrow}
\right],
\nonumber\\
     \dot {\L}^{\downarrow\uparrow}_{00}&=&
\frac{1}{m}\P_{00}^{\downarrow\uparrow}-m\,\omega^2\R^{\downarrow\uparrow}_{00}
+4\sqrt3\kappa_0 R_{20}^{\rm eq}\,\R^{\downarrow\uparrow}_{20}
\nonumber\\
&&+\frac{1}{2\sqrt{3}}\hbar^2 
\left[\left( \chi-\frac{\bar\chi}{3} \right)I_1-\frac{9}{4}\chi I_2\right]
\left[\left(\frac2{\A_2}-\frac1{\A_1}\right)\R_{00}^{\downarrow\uparrow}+
\sqrt2\left(\frac1{\A_2}+\frac1{\A_1}\right)\R_{20}^{\downarrow\uparrow}
\right],
\nonumber\\
     \dot {\R}^{\downarrow\uparrow}_{00}&=&
\frac{2}{m}\L_{00}^{\downarrow\uparrow},
\nonumber\\
     \dot {\P}^{\downarrow\uparrow}_{00}&=&
-2m\,\omega^2\L^{\downarrow\uparrow}_{00}
+8\sqrt3\kappa_0 R_{20}^{\rm eq}\,\L^{\downarrow\uparrow}_{20}
+\frac{\sqrt{3}}{2}\hbar^2 
\chi I_2
\left[\left(\frac2{\A_2}-\frac1{\A_1}\right)\L_{00}^{\downarrow\uparrow}+
\sqrt2\left(\frac1{\A_2}+\frac1{\A_1}\right)\L_{20}^{\downarrow\uparrow}
\right],
 \nonumber\\
     \dot{ {\tilde\R}}_{21} &=& -\frac{1}{\hbar}\left(\frac{16}{5} \alpha\kappa_0 \K_4
     +\Delta_0(r') -\frac{3}{8}\hbar^2 \chi\kappa_0(r') \right)\R^+_{21}, 
 \nonumber\\
     \dot{ {\tilde\P}}_{21} &=& -\frac{1}{\hbar}\Delta_0(r') \P^+_{21} + 
 6 \hbar\alpha\kappa_0 \K_0{\cal R}^+_{21}, 
\nonumber\\
     \dot {{\tilde\L}}_{21} &=& -\frac{1}{\hbar}\Delta_0(r') {\L}^-_{21},
\nonumber\\
     \dot {{\tilde\L}}_{11} &=& -\frac{1}{\hbar}\Delta_0(r') \L^-_{11},
\end{eqnarray}
where $
{\cal A}_1=\sqrt2\, R_{20}^{\rm eq}-R_{00}^{\rm eq}=\frac{Q_{00}}{\sqrt3}\left(1+\frac{4}{3}\delta\right),\quad
{\cal A}_2= R_{20}^{\rm eq}/\sqrt2+R_{00}^{\rm eq}=-\frac{Q_{00}}{\sqrt3}\left(1-\frac{2}{3}\delta\right)
$,
$\quad\displaystyle a_{-1} = a_1 = R_0\left( \frac{1-(2/3)\delta}{1+(4/3)\delta} \right)^{1/6}$ and
$\displaystyle a_0 = R_0\left( \frac{1-(2/3)\delta}{1+(4/3)\delta} \right)^{-1/3}$ 
are semiaxes of ellipsoid by which the shape of nucleus is approximated, $\delta$ -- deformation parameter,
$R_0=1.2A^{1/3}$~fm.
\begin{eqnarray}\label{Int_ss} \nonumber
I_1=\frac{\pi}{4}\int\limits_{0}^{\infty}dr\, r^4\left(\frac{\partial n^+(r)}{\partial r}\right)^2,
\
I_2=\frac{\pi}{4}\int\limits_{0}^{\infty}dr\, r^2 n^+(r)^2,\quad 
n^+(r)=n_p^{+}+n_n^{+}= \frac{n_0}{1+{\rm e}^{\frac{r-R_0}{a}}}.
\end{eqnarray}
$\K_0=\int d(\br,\bp) \kappa_0(\br,\bp), \, \K_4=\int d(\br,\bp) r^4\kappa_0(\br,\bp)$.
The functions $\kappa_0(r')$, $\Delta_0(r')$, $I_{rp}^{\kappa\Delta}(r')$ and 
$I_{pp}^{\kappa\Delta}(r')$
are discussed in the next section and are demonstrated in Appendix D.
Deriving these equations we neglected double Poisson brackets containing $\kappa$ or $\Delta$,
which are the quantum corrections to pair correlations.
The isoscalar set of equations is easily obtained from (\ref{iv}) by 
taking $\alpha=1$, replacing $\bar\chi \to -\bar\chi$
and putting the marks "bar" above all variables.

\section{Results of calculations}

The set of equations (\ref{iv}) coincides with the set of equations (27) of the paper \cite{BaMoPRC}
in the limit of zero pairing, i.e. if to omit the last four equations and
to neglect the  contributions from pairing in the dynamical
equations for the variables ${\L}^-_{21},\,{\L}^-_{11},$ and ${\P}^+_{21}$ .
On the other hand, the dynamical equations for ${\tilde \P}_{21}$ and ${\tilde \R}_{21}$
and the contribution from pairing in the dynamical equation for ${\P}_{21}^+$ are
exactly the same as the ones in the paper \cite{Urban}. Only the dynamical equations for
${\tilde \L}_{21},\,$ ${\tilde \L}_{11}$ and the contributions from pairing in
dynamical equations for ${\L}^-_{21},\,$ ${\L}^-_{11}$ are completely new.

Imposing the time evolution via $\di{e^{iEt/\hbar}}$ for all variables
one transforms (\ref{iv}) into a set of algebraic equations. 
It contains 23 equations. 
To find the eigenvalues we construct the 23x23
determinant and seek (numerically) for its zeros. 
We find seven roots with exactly E=0 and 16 roots which are non zero: 
eight positive ones and eight negative ones (situation is exactly
same as with RPA; see \cite{Ann} for connection of WFM and RPA). In this paper we consider
only the two lowest roots corresponding to the orbital and spin scissors. The qualitative picture
of high lying modes remains practically without any changes in comparison with 
\cite{BaMoPRC}.

 Seven integrals of motion corresponding to Goldstone modes (zero roots)
can be found analytically. They are written out in the Appendix C.
The interpretation of some of them has been found in \cite{BaMoPRC}, whereas the interpretation 
of the remaining ones seems not to be obvious.

\subsection{Choice of parameters}
 
$\bullet$ Following our previous publications \cite{BaSc,Ann} we take for the isoscalar strength 
constant of the quadrupole-quadrupole residual interaction $\kappa_0$ the self consistent 
value \cite{BrMt} $\kappa_0=-\frac{m\bar\omega^2}{4Q_{00}}$ with
$Q_{00}=A\frac{3}{5}R^2$, $R=r_0A^{1/3}$,
$r_0=1.2$ fm, $\bar\omega^2=\omega_0^2/
[(1+\frac{4}{3}\delta)^{2/3}(1-\frac{2}{3}\delta)^{1/3}]$,
$\hbar\omega_0=41/A^{1/3}$ MeV. 

\begin{figure}[h]
\centering\includegraphics[width=8cm]{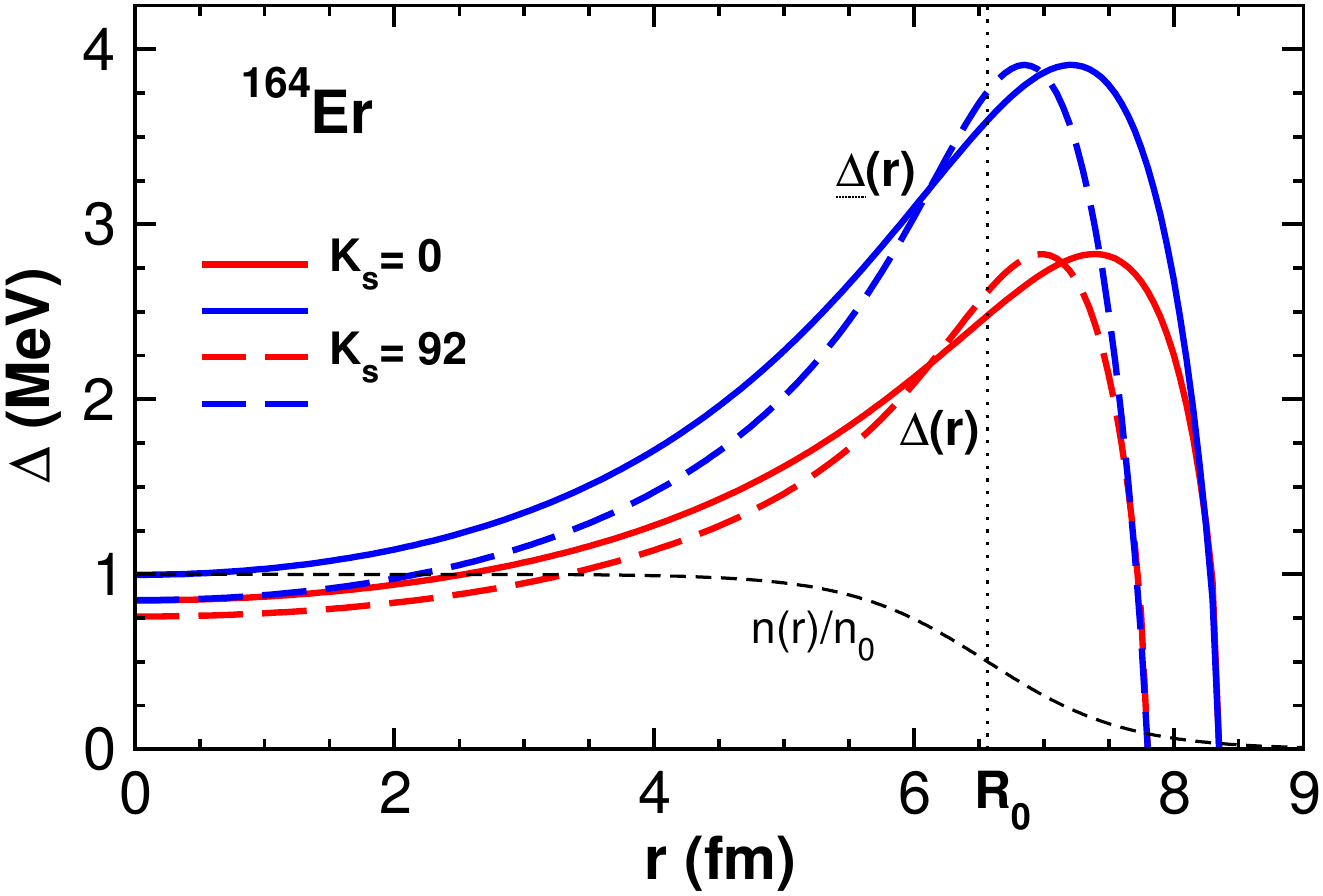}
\caption{The pair field (gap) $\Delta_0(r)$, the function 
$\underline{\Delta}=|V_0|I_{pp}^{\kappa\Delta}(r)$ and the nuclear density $n(r)$ 
as the functions of radius $r$. The solid lines -- calculations without the spin-spin 
interaction $V_{ss}$, the dashed lines -- $V_{ss}$ is included.}
\label{fig1}\end{figure}

$\bullet$
The equations (\ref{iv}) contain the functions 
$\Delta_0(r')\equiv\Delta_{eq}(r',p_F(r'))$, 
$I_{rp}^{\kappa\Delta}(r')\equiv I_{rp}^{\kappa\Delta}(r',p_F(r'))$, 
$I_{pp}^{\kappa\Delta}(r')\equiv I_{pp}^{\kappa\Delta}(r',p_F(r'))$ 
and $\kappa_0(r')\equiv\kappa(r',r')$ 
depending on the radius $r'$  and the local Fermi momentum $p_F(r')$ 
(see Fig.~\ref{fig1} ).
The value of $r'$ is not fixed by the theory and can be used 
as the fitting parameter. We have found in our previous paper \cite{Urban} that the best agreement
of calculated results with experimental data is achieved at the point $r'$ where the function
$I_{pp}^{\kappa\Delta}(r',p_F(r'))$ has its maximum. Nevertheless, to get rid off the fitting 
parameter, we use the averaged values of these functions:
$\bar\Delta_0=\int d\br\, n_0(\br)\Delta_0(r,p_F(r))/A$, etc.
The gap $\Delta(r,p_F(r))$, as well as the integrals $I^{\kappa\Delta}_{pp}(r,p_F(r))$, $\K_4$
and $\K_0$,
were calculated with the help of the semiclassical 
formulae for $\kappa(\br,\bp)$ and $\Delta(\br,\bp)$ (see
Appendix D), a Gaussian being
used for the pairing interaction with $r_p=1.9$ fm and $V_0=25$
MeV \cite{Ring}. Those values reproduce usual nuclear pairing gaps.

$\bullet$
The used spin-spin interaction is repulsive, the values of its
strength constants being taken from the paper \cite{Moya}, where the 
notation $\chi=K_s/A,\,\bar\chi=q\chi$ was introduced.
The constants were extracted by the authors of~\cite{Moya} 
from Skyrme forces following the standard procedure, the residual interaction 
being defined in terms of second derivatives of the Hamiltonian density 
$H(\rho)$ with respect to the one-body densities~$\rho$.
Different variants of Skyrme forces produce different strength constants of 
spin-spin interaction. The most consistent results are obtained with 
SG1, SG2 \cite{Giai} and Sk3 \cite{Floc} forces. 
To compare theoretical results with experiment the authors of \cite{Moya} preferred 
to use the force SG2. Nevertheless they have noticed that "As is well known, the energy 
splitting of the HF states around the Fermi level is too large. This has an effect on the
spin M1 distributions that can be roughly compensated by reducing the $K_s$ value". According 
to this remark they changed the original self-consistent SG2 parameters from $K_s=88$ MeV, 
$q=-0.95$ to $K_s=50$ MeV, $q=-1$. It was  found that this modified set of parameters
gives better agreement with experiment for some nuclei in the description of spin-flip
resonance. So we will use $K_s=50$ MeV and $q=-1$.

\begin{figure}[h]
\centering\includegraphics[width=8cm]{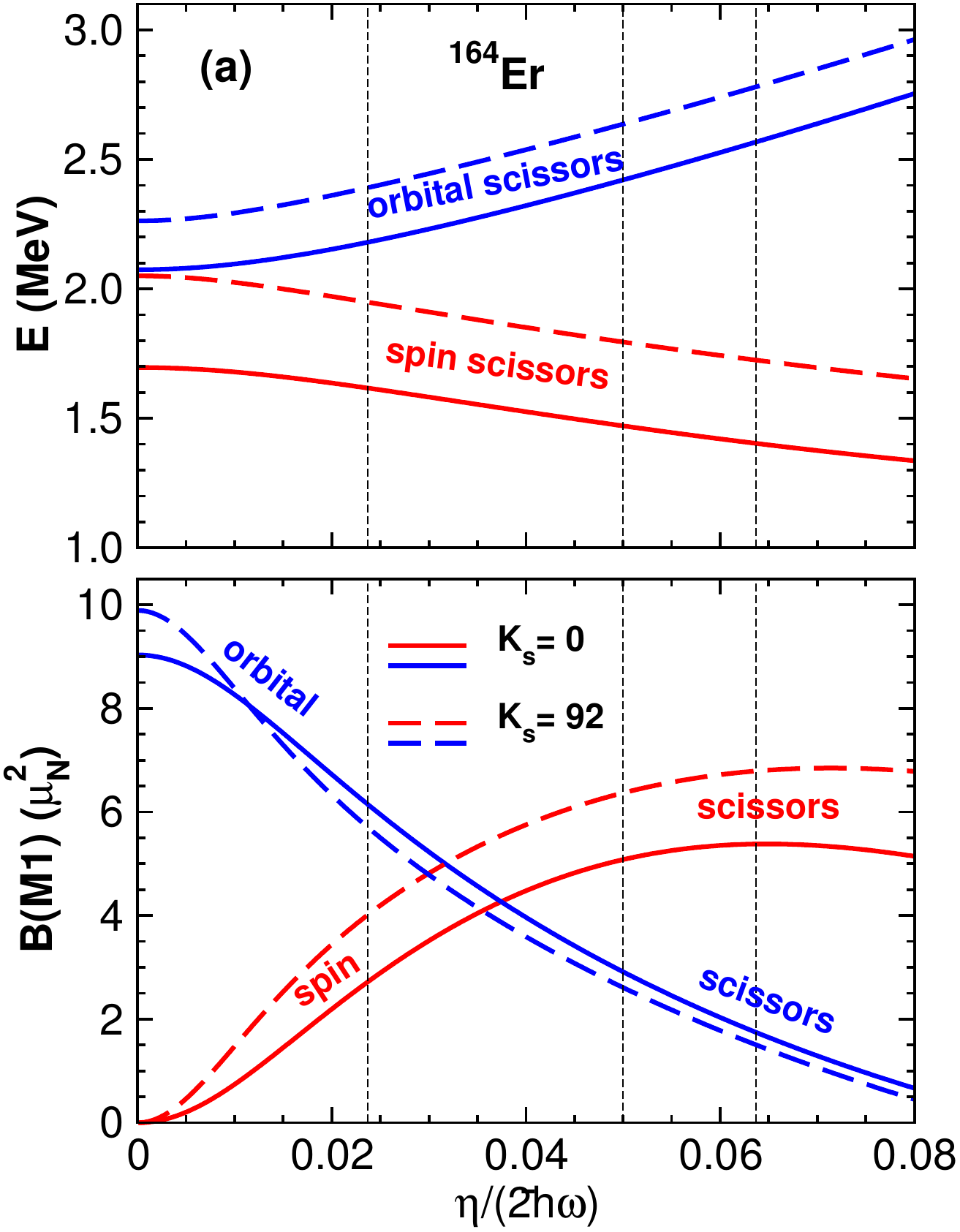}\includegraphics[width=8cm]{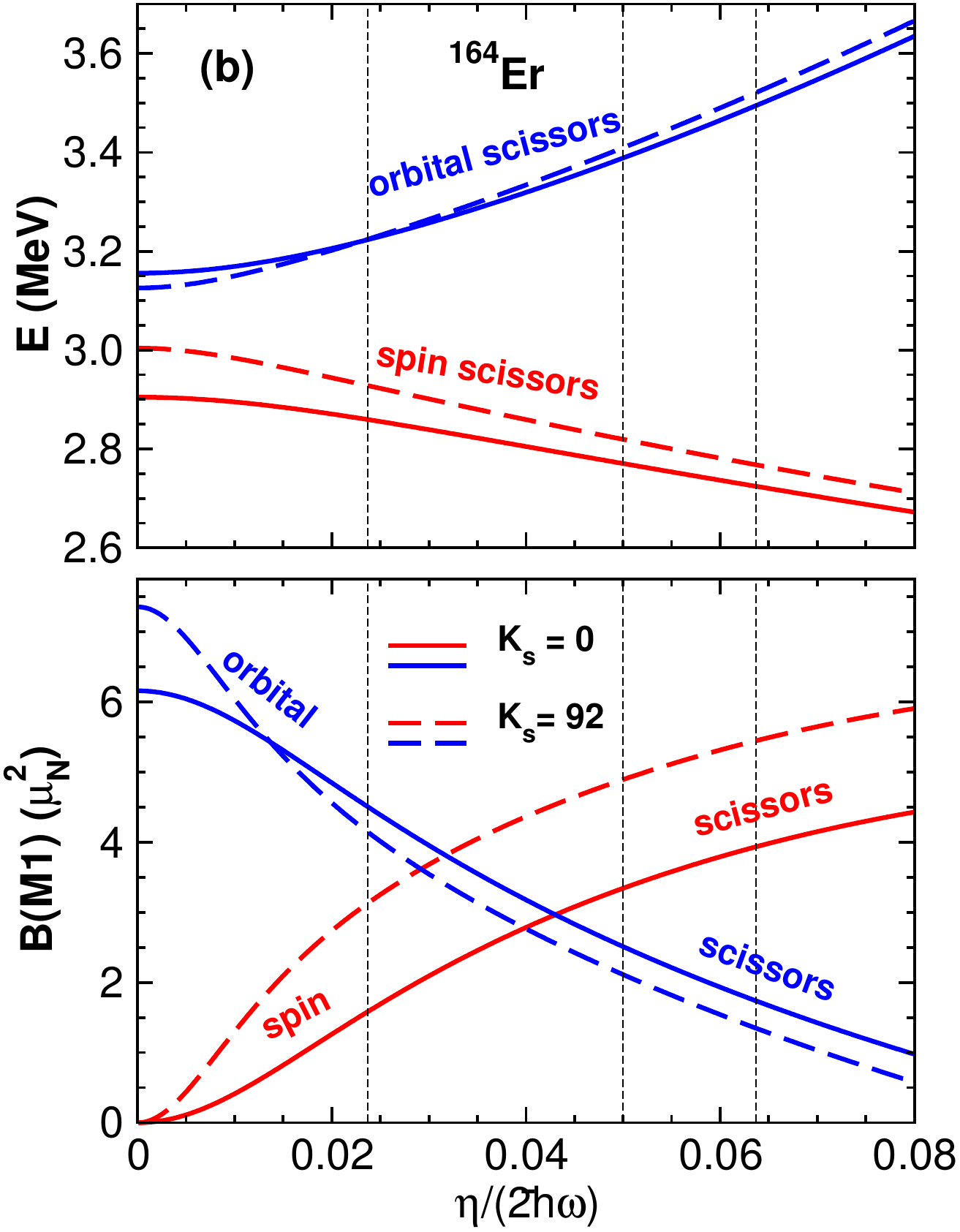}
\caption{The energies $E$ and $B(M1)$-factors as a functions of the spin-orbital strength 
constant $\eta$. 
Left panel: solid lines -- without the spin-spin interaction $V_{ss}$, dashed lines -- $V_{ss}$ is included.
Right panel: The same as in left panel with pair correlations included.}
\label{fig2}\end{figure}

$\bullet$
Our calculations without pairing \cite{BaMoPRC} have shown that the results are
strongly dependent on the values  of 
the strength constants of the spin-spin interaction. The natural question arises: how sensitive are they
to the strength of the spin-orbital potential? The results of the demonstrative calculations 
are shown in~Fig.~\ref{fig2}.

The $M1$ strengths were computed using effective spin giromagnetic factors 
$g_s^{eff}=0.7g_s^{free}$. One observes a rather strong dependence of the results
on the value of $\eta$: the splitting $\Delta E$ and the $M1$ strength of the spin scissors grow with 
increasing $\eta$, the $B(M1)$ of the orbital scissors being decreased.
At some critical point $\eta_c$ the $M1$ strength of the spin scissors becomes bigger than that of the
orbital scissors. The inclusion of the spin-spin interaction does not change the qualitative picture,
as well as the inclusion of pair correlations (see Fig.~\ref{fig2}).

What value of $\eta$ to use?
 Accidentally, the choice of $\eta$ in our previous papers~\cite{BaMo,BaMoPRC} was not very realistic. 
The main purpose  of the first paper was the introduction 
of spin degrees of freedom into the WFM method, and the aim of the second paper
was to study the influence of spin-spin forces on both scissors -- we did not worry much about the comparison
with experiment. Now, both preliminary aims being achieved, one can think about the agreement with
experimental data, therefore the precise choice of the model parameters becomes important. Of course,
we could try to choose $\eta$ according to the standard requirement of the best agreement with experiment.
 However, in reality we are not absolutely free in our choice.
It turns out that we are already restricted by the other constraints. As a matter of fact
we work with the Nilsson potential, parameters of which are very well known. Really, the mean field of our
model (\ref{Ham}) is the deformed harmonic oscillator with the spin-orbit potential, the Nilsson 
$\ell^2$ term being neglected because it generates the fourth order moments and, anyway, they are probably not of great importance. 
In the original paper~\cite{Nils} Nilsson took the spin-orbit strength constant
 $\kappa_{Nils}=0.05$ for rare earth nuclei. 
Later the best value of $\kappa_{Nils}$
for rare earth nuclei was established~\cite{Ring} to be $0.0637$. For actinides there were established 
different values of $\kappa_{Nils}$ for neutrons ($0.0635$) and protons ($0.0577$).
The numbers $\kappa_{Nils}=0.0637$, $\kappa_{Nils}=0.05$ 
and $\kappa_{Nils}=0.024$ (corresponding to $\eta=0.36$ used in our previous calculations~\cite{BaMo,BaMoPRC}) 
are marked on~Figs~\ref{fig2}, \ref{figL} by 
the dotted vertical lines. Of course we will use the
conventional \cite{Ring} parameters of the Nilsson potential and
from now on we will speak only about the Nilsson \cite{Nils} spin-orbital strength parameter 
$\kappa_{Nils}$, which is connected with~$\eta$ by the relation $\eta=2\hbar\omega\kappa_{Nils}$.

\subsection{Discussion and interpretation of results}

The energies and excitation probabilities of orbital and spin scissors modes obtained by the 
solution of the isovector set of equations (\ref{iv}) are displayed in the Table 1.

\begin{table}[h!]\label{tab1}
\caption{Scissors modes energies $E_{\rm sc}$ and
transition probabilities $B(M1)_{\rm sc}$.}
\begin{tabular}{c|c|c|c|c|c|c|c}
\hline
\multicolumn{2}{c|}{ $^{164}$Er }    &
\multicolumn{3}{c|}{ $E_{\rm sc}$, MeV} & \multicolumn{3}{c}{ $B(M1)_{\rm sc},\ \mu_N^2$}    \\
\cline{3-8}
\multicolumn{2}{c|}{  }    &
$K_{\rm s}=0$ & $K_{\rm s}=50$ & $K_{\rm s}=92$ & 
$K_{\rm s}=0$ & $K_{\rm s}=50$ & $K_{\rm s}=92$ \\
\hline
 spin     & $\bar\Delta_0=0$     & 1.40 & 1.60 & 1.73 & 5.38 & 6.23 & 6.79  \\[-2mm]
 scissors & $\bar\Delta_0\neq 0$ & 2.72 & 2.75 & 2.77 & 3.93 & 4.79 & 5.44  \\ \hline
 orbital  & $\bar\Delta_0=0$     & 2.57 & 2.69 & 2.78 & 1.74 & 1.59 & 1.50  \\[-2mm]
 scissors & $\bar\Delta_0\neq 0$ & 3.49 & 3.51 & 3.52 & 1.74 & 1.51 & 1.35  \\
\hline
\end{tabular}
\end{table}

There are results of calculations with three values of the spin-spin strength constant and two
values of $\bar \Delta_0$. As it was expected the energies of both scissors increased approximately
by 1 Mev after inclusion of pairing. The behaviour of transition probabilities turned out less 
predictable. The $B(M1)$ value of the spin scissors decreased approximately by 1.5 $\mu_N^2$, whereas
$B(M1)$ value of the orbital scissors turned out practically insensitive to the inclusion of
pair correlations.

We can compare the summed $B(M1)_{\Sigma}= B(M1)_{or}+B(M1)_{sp}$ values and the centroid
of both scissors energies
$$E_{cen}=[E_{or}B(M1)_{or}+E_{sp}B(M1)_{sp}]/B(M1)_{\Sigma}$$
with the results of the paper \cite{Urban} where no spin degrees of freedom had been considered and with the experimental data. The respective results
are shown in the Table 2. 

\begin{table}[h!]\label{tab2}
\caption{Scissors modes energy centroid $E_{cen}$ and summarized
transition probabilities $B(M1)_{\Sigma}$. 
The experimental values
of $E_{cen}$, $\delta$, and $B(M1)_{\Sigma}$ are from \cite{Pietr} and
references therein.}
\begin{tabular}{c|c|c|c|c|c|c|c|c|c|c}
\hline
\multicolumn{1}{c|}{ $^{164}$Er }    &
\multicolumn{5}{c|}{ $E_{cen}$, MeV} & \multicolumn{5}{c}{ $B(M1)_{\Sigma},\ \mu_N^2$}    \\
\cline{2-11}
    &
$K_{\rm s}=0$ & $K_{\rm s}=50$& $K_{\rm s}=92$ & \cite{Urban} & exp &
$K_{\rm s}=0$ & $K_{\rm s}=50$& $K_{\rm s}=92$ & \cite{Urban} & exp  \\
\hline
$\bar\Delta_0=0$     & 1.69 & 1.82 & 1.92 & 2.10 &      & 7.13 & 7.82 & 8.29 & 9.26 &       \\[-5mm]
                     &      &      &      &      & 2.90 &      &      &      &      & 1.45  \\[-5mm]
$\bar\Delta_0\neq 0$ & 2.96 & 2.93 & 2.92 & 3.37 &      & 5.67 & 6.30 & 6.79 & 5.62 &       \\
\hline
\end{tabular}
\end{table}

It is seen that the inclusion of spin degrees of freedom in the WFM
method does not change markedly our results (in comparison with previous ones
\cite{Urban}). Of course, the energy changed in the desired direction and now practically 
coincides with the experimental value (especially in the case with spin-spin forces.) However,
the situation with the $B(M1)$ values did not change (and even become worse in the case 
with spin-spin forces). Our hope, that spin degrees of freedom can improve
the situation with the  $B(M1)$ values, did not become true: the theory so far gives four times bigger values of 
$B(M1)$ than the experimental ones, exactly as it was the case in the paper \cite{Urban}.

The result look discouraging. However,
a phenomenon, which was missed in our previous papers and  described in the next
section will save the situation.

\section{Counter-rotating angular momenta of spins up/down
(hidden angular momenta)}

The equilibrium (ground state) orbital angular momentum  of any
nucleus is composed of two equal parts: half of nucleons (protons + neutrons) having spin projection up and other half having 
spin projection down. It is known that 
the huge majority of nuclei have zero angular momentum in the ground state. We will show 
below that as a rule this zero is just the sum of two rather big counter directed angular 
momenta (hidden angular momenta, because they are not manifest in the ground state)
of the above mentioned two parts of any nucleus.  Being connected with the spins of 
nucleons this phenomenon naturally has great influence 
on all nuclear properties connected with the spin, in particular, the spin scissors mode.

Let us analyze the procedure of linearization of the equations of motion for collective variables
(\ref{Varis}). We consider small deviations of the system from  equilibrium, so all variables 
are written as a sum of their equilibrium value plus a small deviation:
$$L(t)=L(eq)+\L(t),\quad \mbox{et al.}$$
Neglecting quadratic deviations one obtains the set of linearized equations for deviations depending on the equilibrium 
values $R^{\tau\varsigma}_{\lambda\mu}(eq)$ and $L^{\tau\varsigma}_{\lambda\mu}(eq)$, which are the input data
of the problem. In the paper \cite{BaMoPRC} we made the following choice:
\begin{eqnarray}
&&R^{+}_{2\pm1}(eq)=R^{+}_{2\pm2}(eq)=0,\quad R^{+}_{20}(eq)\neq0,\quad R^{+}_{00}(eq)\neq0,
\label{eq1}
\\
&&R^{\uparrow\downarrow}_{\lambda\mu}(eq)=R^{\downarrow\uparrow}_{\lambda\mu}(eq)=0,
\label{eq3}
\\
&&L^{\tau\varsigma}_{\lambda\mu}(eq)=0,\quad
R^{-}_{\lambda\mu}(eq)=0.
\label{eq4}
\end{eqnarray}

At first glance, this choice looks quite natural. Really, relations (\ref{eq1}) 
follow from the axial symmetry of nucleus. Relations (\ref{eq3}) are justified by the fact that these quantities should be diagonal in spin at equilibrium. The variables
$L^{\tau\varsigma}_{\lambda\mu}(t)$ contain the momentum $\bp$ in their definition which 
incited us to suppose  zero equilibrium values as well (we will show below that it is not 
true for $L^-_{10}$ because of quantum effects connected with spin). 

The relation $R^{-}_{\lambda\mu}(eq)=0$ follows from the shell model considerations: the 
nucleons with spin projection "up" and "down" are sitting in pairs on the same levels, therefore all
average properties of the "spin up" part of nucleus must be identical to that of the "spin down" part.
However, the careful analysis shows that being undoubtedly true for variables
$R^{\uparrow\uparrow}_{\lambda\mu},\,R^{\downarrow\downarrow}_{\lambda\mu}$ this statement turns out
erroneous for variables $L^{\uparrow\uparrow}_{10},\,L^{\downarrow\downarrow}_{10}$. Let us demonstrate it. 
By definition
\begin{eqnarray}
L^{ss'}_{\lambda\mu}(t)=
\int\! d^3r\,\int\! \frac{d^3p}{(2\pi\hbar)^{3}} \{r\otimes p\}_{\lambda\mu}
f^{ss'}(\br,\bp,t)=
\int\! d^3r \{r\otimes J^{ss'}\}_{\lambda\mu},
\label{LJ}
\end{eqnarray}
where 
\begin{eqnarray}
J^{ss'}_i(\br,t)=\int\! \frac{d^3p}{(2\pi\hbar)^{3}} p_{i}f^{ss'}(\br,\bp,t)=
\int\! \frac{d^3p}{(2\pi\hbar)^{3}} p_{i}\int\!d^3q\exp(-\frac{i}{\hbar}\bp\cdot\bq)
\rho(\br+\frac{\bq}{2},s;\br-\frac{\bq}{2},s';t)
\label{J}
\end{eqnarray}
is the i-th component of the nuclear current. In the last relation the definition \cite{Ring} of Wigner function is used. 
Performing the integration over $\bp$ one finds:
\begin{eqnarray}
J^{ss'}_i(\br,t)=i\hbar\int\!d^3q[\frac{\partial}{\partial q_i}\delta(\bq)]
\rho(\br+\frac{\bq}{2},s;\br-\frac{\bq}{2},s';t)
\nonumber\\
=-i\hbar\int\!d^3q\delta(\bq)\frac{\partial}{\partial q_i}
\rho(\br+\frac{\bq}{2},s;\br-\frac{\bq}{2},s';t)
\nonumber\\
=-\frac{i\hbar}{2}[(\nabla_{1i}-\nabla_{2i})\rho(\br_1,s;\br_2,s';t)]_{\br_1=\br_2=\br},
\label{J1}
\end{eqnarray}
where $\br_1=\br+\frac{\bq}{2},\, \br_2=\br-\frac{\bq}{2}$. The density matrix of the ground state
nucleus is defined \cite{Ring} as
\begin{equation}
\rho(\br_1,s;\br_2,s';t)=\sum_{\nu}v^2_{\nu}\phi_{\nu}(\br_1s)\phi^*_{\nu}(\br_2s'),
\label{rho}
\end{equation}
where $v^2_{\nu}$ are occupation numbers and $\phi_{\nu}$ are single particle wave functions. For the sake
of simplicity we will consider the case of spherical symmetry. Then $\nu=nljm$ and
\begin{eqnarray}
\phi_{nljm}(\br,s)=\R_{nlj}(r)\sum_{\Lambda,\sigma}C^{jm}_{l\Lambda,\frac{1}{2}\sigma}
Y_{l\Lambda}(\theta,\phi)\chi_{\frac{1}{2}\sigma}(s),
\label{phi}
\end{eqnarray}
\begin{eqnarray}
J^{ss'}_i(\br)=-\frac{i\hbar}{2}\sum_{\nu}v^2_{\nu}
[\nabla_i\phi_{\nu}(\br,s)\cdot\phi^*_{\nu}(\br,s')
-\phi_{\nu}(\br,s)\cdot\nabla_i\phi^*_{\nu}(\br,s')]
\label{Jdef}
\\
=-\frac{i\hbar}{2}\sum_{nljm}v^2_{nljm}\R^2_{nlj}\sum_{\Lambda,\sigma,\Lambda',\sigma'}
C^{jm}_{l\Lambda,\frac{1}{2}\sigma}C^{jm}_{l\Lambda',\frac{1}{2}\sigma'}
[Y^*_{l\Lambda'}\nabla_iY_{l\Lambda}-Y_{l\Lambda}\nabla_iY^*_{l\Lambda'}]
\chi_{\frac{1}{2}\sigma}(s)\chi_{\frac{1}{2}\sigma'}(s').
\label{Jphi}
\end{eqnarray}
Inserting this expression into (\ref{LJ}) one finds:
\begin{eqnarray}
L^{ss'}_{10}(eq)=
-\frac{i\hbar}{2}\sum_{nljm}v^2_{nljm}\sum_{\Lambda\sigma,\Lambda'\sigma'}
C^{jm}_{l\Lambda,\frac{1}{2}\sigma}C^{jm}_{l\Lambda',\frac{1}{2}\sigma'}
\chi_{\frac{1}{2}\sigma}(s)\chi_{\frac{1}{2}\sigma'}(s')
\nonumber\\
\int\! d^3r\,\R^2_{nlj}[Y^*_{l\Lambda'}\{r\otimes\nabla\}_{10}Y_{l\Lambda}-Y_{l\Lambda}\{r\otimes\nabla\}_{10}Y^*_{l\Lambda'}]
\nonumber\\
=\frac{i}{2\sqrt2}\sum_{nljm}v^2_{nljm}\sum_{\Lambda\sigma,\Lambda'\sigma'}
C^{jm}_{l\Lambda,\frac{1}{2}\sigma}C^{jm}_{l\Lambda',\frac{1}{2}\sigma'}
\chi_{\frac{1}{2}\sigma}(s)\chi_{\frac{1}{2}\sigma'}(s')
\int\! d^3r\,\R^2_{nlj}[Y^*_{l\Lambda'}\hat l_0Y_{l\Lambda}-Y_{l\Lambda}\hat l_0Y^*_{l\Lambda'}]
\nonumber\\
=\frac{i}{2\sqrt2}\sum_{nljm}v^2_{nljm}\sum_{\Lambda\sigma,\Lambda'\sigma'}
C^{jm}_{l\Lambda,\frac{1}{2}\sigma}C^{jm}_{l\Lambda',\frac{1}{2}\sigma'}
\chi_{\frac{1}{2}\sigma}(s)\chi_{\frac{1}{2}\sigma'}(s')(\Lambda+\Lambda')\delta_{\Lambda,\Lambda'}
\nonumber\\
=\frac{i}{\sqrt2}\sum_{nljm}v^2_{nljm}\sum_{\Lambda\sigma}\Lambda\,
\left(C^{jm}_{l\Lambda,\frac{1}{2}\sigma}\right)^2
\chi_{\frac{1}{2}\sigma}(s)\chi_{\frac{1}{2}\sigma}(s').
\label{L10}
\end{eqnarray}
Here the definition $\hat l_{\mu}=-\hbar\sqrt{2}\{r\otimes\nabla\}_{1\mu}$, formula 
$\hat l_0Y_{l\Lambda}=\Lambda Y_{l\Lambda}$ and normalization of functions $\R_{nlj}$ were used.
Remembering the definition of the spin function $\chi_{\frac{1}{2}\sigma}(s)=\delta_{\sigma,s}$
we get finally:
\begin{equation}
L^{ss'}_{10}(eq)=
\frac{i}{\sqrt2}\sum_{nljm}v^2_{nljm}\sum_{\Lambda}\Lambda\,
\left(C^{jm}_{l\Lambda,\frac{1}{2}s}\right)^2\delta_{s,s'}=
\delta_{s,s'}\frac{i}{\sqrt2}\sum_{nljm}v^2_{nljm}
\left(C^{jm}_{lm-s,\frac{1}{2}s}\right)^2(m-s).
\label{L10f}
\end{equation}
Now, with the help of analytic expressions for Clebsh-Gordan coefficients one obtains the final  
expressions
\begin{eqnarray}
L^{\uparrow\uparrow}_{10}(eq)
=\frac{i}{\sqrt2}\sum_{nl}\left[
\sum_{m=-\left(l+\frac{1}{2}\right)}^{l+\frac{1}{2}}v^2_{nlj^+m}\frac{l+\frac{1}{2}+m}{2l+1} +
\sum_{m=-\left(l-\frac{1}{2}\right)}^{l-\frac{1}{2}}v^2_{nlj^-m}\frac{l+\frac{1}{2}-m}{2l+1}
\right]\left(m-\frac{1}{2}\right),
\label{L10up}
\end{eqnarray}
\begin{eqnarray}
L^{\downarrow\downarrow}_{10}(eq)
=\frac{i}{\sqrt2}\sum_{nl}\left[
\sum_{m=-\left(l+\frac{1}{2}\right)}^{l+\frac{1}{2}}v^2_{nlj^+m}\frac{l+\frac{1}{2}-m}{2l+1} +
\sum_{m=-\left(l-\frac{1}{2}\right)}^{l-\frac{1}{2}}v^2_{nlj^-m}\frac{l+\frac{1}{2}+m}{2l+1}
\right]\left(m+\frac{1}{2}\right),
\label{L10d}
\end{eqnarray}
where the notation $j^{\pm}=l\pm\frac{1}{2}$ is introduced. Replacing in (\ref{L10up}) $m$
by $-m$ we find that
\begin{eqnarray}
L^{\uparrow\uparrow}_{10}(eq)=-L^{\downarrow\downarrow}_{10}(eq).
\label{L10upd}
\end{eqnarray}
By definition (\ref{Varis}) 
$L^{\pm}_{10}(eq)=L^{\uparrow\uparrow}_{10}(eq) \pm L^{\downarrow\downarrow}_{10}(eq)$.
Combining linearly (\ref{L10up}) and (\ref{L10d}) one finds:
\begin{eqnarray}
L^+_{10}(eq)
=\frac{i}{\sqrt2}\sum_{nl}\left[
\sum_{m=-\left(l+\frac{1}{2}\right)}^{l+\frac{1}{2}}v^2_{nlj^+m}\frac{2l}{2l+1}m+
\sum_{m=-\left(l-\frac{1}{2}\right)}^{l-\frac{1}{2}}v^2_{nlj^-m}\frac{2l+2}{2l+1}m
\right],
\label{L10+}
\end{eqnarray}
\begin{eqnarray}
L^-_{10}(eq)=\frac{i}{\sqrt2}\sum_{nl}\left[
\sum_{m=-\left(l+\frac{1}{2}\right)}^{l+\frac{1}{2}}v^2_{nlj^+m}\frac{2m^2-l-\frac{1}{2}}{2l+1}-
\sum_{m=-\left(l-\frac{1}{2}\right)}^{l-\frac{1}{2}}v^2_{nlj^-m}\frac{2m^2+l+\frac{1}{2}}{2l+1}
\right].
\label{L10-}
\end{eqnarray}
These formulas are valid for spherical nuclei. However, with the scissors and spin-scissors modes, we are considering deformed nuclei. 
For the sake of the discussion,
let us consider the case of infinitesimally small deformation, when one can continue to use formulae~(\ref{L10+}, \ref{L10-}). 
Now only levels with quantum numbers $\pm m$ are degenerate.
According to, for example, the Nilsson scheme~\cite{Nils} nucleons will occupy pairwise 
precisely those levels which leads to the zero 
value of~$L^+_{10}(eq)$.

What about~$L^-_{10}(eq)$? It only enters (27) in the equation for $\dot {\P}^-_{21}$. Let us analyze the structure of formula (\ref{L10-}) considering for 
the sake of simplicity the case without pairing. Two sums over $m$ (let us note them $\Sigma_1$
and $\Sigma_2$) represent two spin-orbital partners: in the first sum the summation goes over
levels of the lower partner ($j=l+\frac{1}{2}$) and in the second sum -- over levels of the
higher partner ($j=l-\frac{1}{2}$). The values of both sums depend naturally on the values of
occupation numbers $n_{nljm}= 0,1$. There are three possibilities. The first one is trivial: if all
levels of both spin-orbital partners are disposed above the Fermi surface, then the respective
occupation numbers $n_{nljm}=0$ and both sums are equal to zero identically. The second 
possibility: all levels of both spin-orbital partners are disposed below the Fermi surface.
Then all respective occupation numbers $n_{nlj^+m}=n_{nlj^-m}=1$. The elementary analytical
calculation (for arbitrary $l$) shows that in this case the two sums in (44) exactly compensate 
each other,
i.e. $\Sigma_1+\Sigma_2=0$. The most interesting is the third possibility, when one part of
levels of two spin-orbital partners is disposed below the Fermi surface and another part is
disposed above it. In this case the compensation does not happen and one gets 
$\Sigma_1+\Sigma_2\neq 0$ what leads to $L^-_{10}(eq)\neq 0$. In the case of pairing, things 
are not so sharply separated and $L_{10}^-(eq)$ has always a finite value. However, the modifications with respect to mean field are very small.

Let us illustrate the above analysis by the example of $^{164}$Er (protons). Its deformation is
$\delta=0.25$ $(\epsilon=0.26)$ and Z=68. Looking on the Nilsson scheme (for example, Fig.1.5 of
\cite{Solov}
or Fig. 2.21c of \cite{Ring}) one easily finds, that only three pairs of spin-orbital partners 
give a nonzero contribution to $L^-_{10}(eq)$. They are: $N=4, d_{5/2}-d_{3/2}$ (two 
levels of $d_{5/2}$ are below the Fermi surface, all the rest -- above); 
$N=4, g_{9/2}-g_{7/2}$ (one level of $g_{7/2}$ is above the Fermi surface, all the rest -- below);
$N=5, h_{11/2}-h_{9/2}$ (four levels of $h_{11/2}$ are below the Fermi surface, all the rest 
-- above).
It is possible to make the crude evaluation of $L^-_{10}(eq)$ using the quantum numbers
indicated on Fig.1.5 of \cite{Solov} or Fig. 2.21c of \cite{Ring}). The result turns out rather
close to the exact one,
computed with the help of formulas (\ref{LJ},\ref{Jdef}) and Nilsson wave functions.
 The influence of pair correlations is very small.

\begin{figure}[h]
\centering\includegraphics[width=8cm]{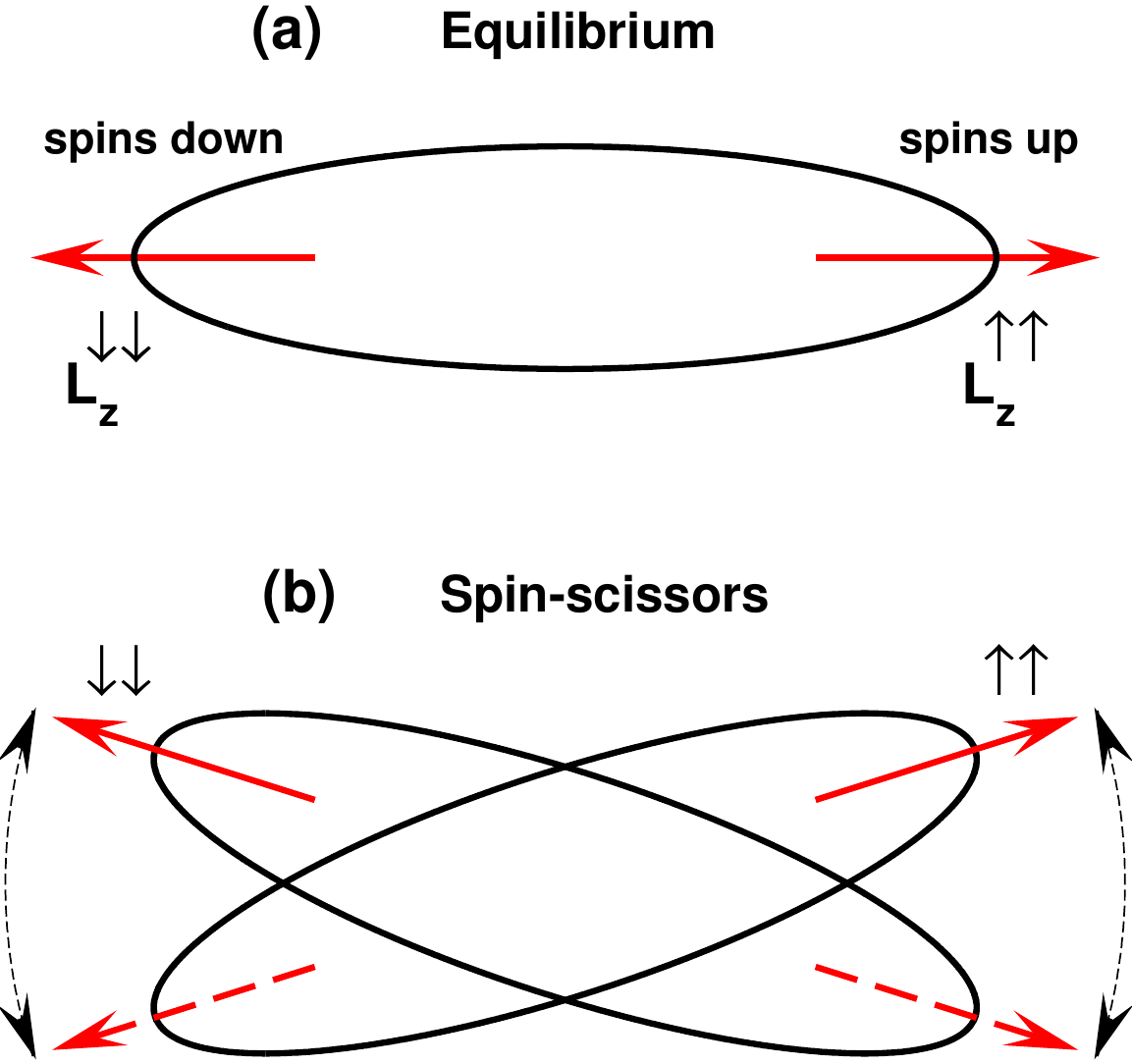}
\caption{(a) Protons with spins $\uparrow$ (up) and $\downarrow$ (down) having nonzero orbital
angular momenta at equilibrium. (b) Protons from Fig.(a) vibrating against one-another.}
\label{figSch}\end{figure}

Indeed, from the definitions (\ref{LJ}) and (\ref{L10}) one can see that $L^{ss}_{10}(eq)$ is just 
the average value of the z-component of the orbital angular momentum of nucleons with the 
spin projection $s$~($\frac{1}{2}$~or~$-\frac{1}{2}$). So,  
the ground state nucleus consists 
of two equal parts having nonzero angular momenta with opposite directions, which compensate 
each other resulting in the zero total angular momentum. 
This is graphically depicted in Fig.~\ref{figSch}(a). 
 
On the other hand, when the opposite angular momenta become tilted, one excites the system and the opposite 
angular momenta are vibrating with a tilting angle, see Fig.~\ref{figSch}(b). 
Actually the two opposite angular momenta are oscillating, one in the opposite sense of the other. 
It is rather obvious from Fig.~\ref{fig0} that these tilted vibrations 
happen separately in each of the neutron and proton lobes. 
These spin-up against spin-down motions certainly influence the 
excitation of the spin scissors mode. 
So, classically speaking
the proton and neutron parts of the ground state nucleus 
consist each of two identical gyroscopes rotating in opposite directions. One knows that it 
is very difficult to deviate gyroscope from an equilibrium. So one can expect, 
that the probability to force two gyroscopes to oscillate as scissors (spin scissors) 
should be small. This picture is confirmed in the next section.

\section{Results of calculations continued}

\begin{figure}[h!]
\centering\includegraphics[width=8cm]{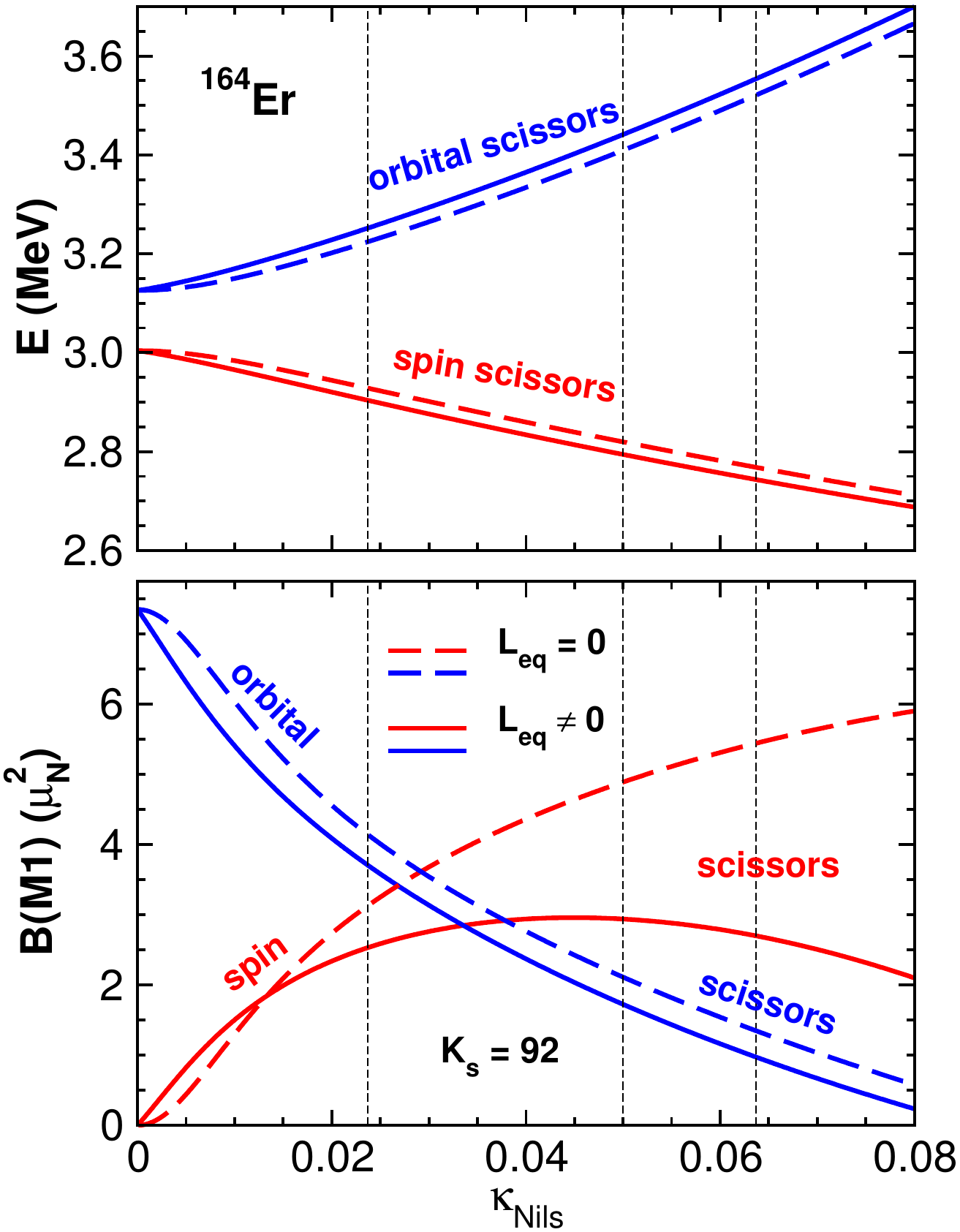}
\caption{The energies $E$ and $B(M1)$-factors as a functions of the spin-orbital strength constant 
$\kappa_{Nils}$. The dashed lines -- calculations without 
$L_{10}^-(eq)$, the solid lines -- $L_{10}^-(eq)$ are taken into account. $V_{ss}$ and pairing 
are included.}
\label{figL}\end{figure}

We made the calculations taking into account the non zero 
value of $L_{10}^-(eq)$ (which was computed according to formulas
(\ref{LJ},\ref{Jdef}) and Nilsson wave functions). The results are shown 
on~Fig.~\ref{figL}. 

They demonstrate (in comparison with~Fig.~\ref{fig2}) the strong
influence of the spin-up vs spin-down angular momenta on the spin scissors mode, whose B(M1) value is strongly 
decreasing with increasing $\kappa_{Nils}$. The B(M1) value of the orbital scissors also
is reduced, but not so much, the value of the reduction being practically independent on
$\kappa_{Nils}$. The influence of $L_{10}^-(eq)$ on the energies of both scissors is negligible,
leading to the small increase of their splitting. Now the energy centroid of both scissors and
their summed B(M1) value at $\kappa_{Nils}=0.0637$ are $E_{cen}=2.97$ MeV and 
$B(M1)_{\Sigma}=3.7 \mu_N^2$. The general agreement with experiment becomes considerably 
better (compare with Table 2), though the theoretical value of $B(M1)_{\Sigma}$ still exceeds the
experimental one approximately 2.5 times. However, as we will see, the case of 
$^{164}$Er may imply a quite particular situation (or even a problem with the experimental value).

\begin{figure}[h!]
\centering\includegraphics[width=8cm]{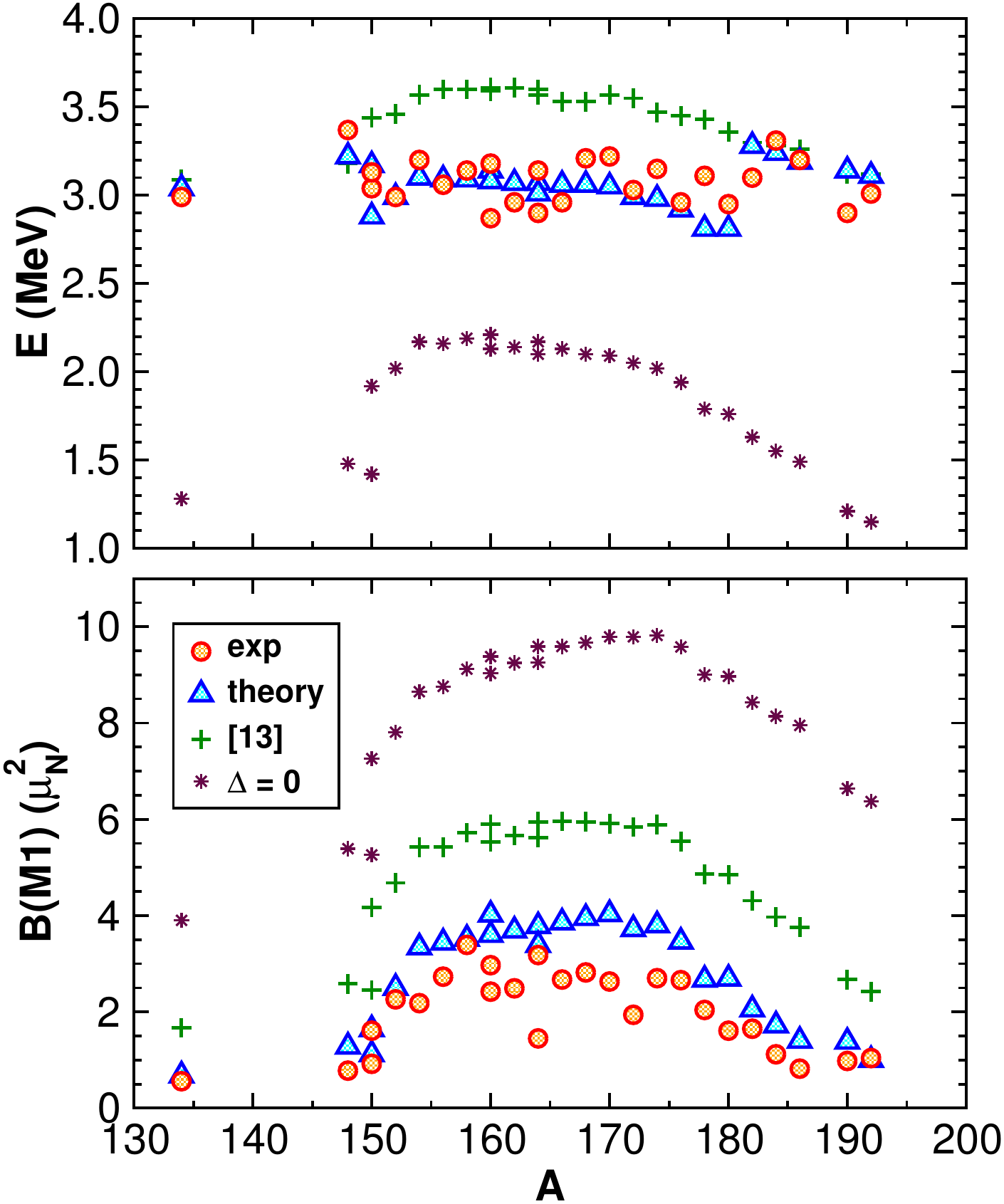}
\caption{The energies $E_{sc}$ and $B(M1)_{sc}$-factors as a function of the mass number A
for nuclei listed in the Table 3.}
\label{figMalov}\end{figure}

The results of systematic calculations for rare-earth nuclei are presented in Tables 3 and
4 and desplayed in Fig.~\ref{figMalov}. Table 3 contains the results for well deformed nuclei with
$\delta \geq 0.18.$
It is easy to see that the overall (general) agreement of theoretical
results with experimental data is substantially improved (in comparison with our previous
calculations \cite{Urban}).

\begin{table}\label{tab3}
\caption{Scissors modes energy centroid $E_{\rm cen}$ and summarized
transition probabilities $B(M1)_{\Sigma}$. Parameters: $\kappa_{Nils}=0.0637$, $V_0=25$
($V_0=27$ for $^{182, 184, 186}$W). The experimental values
of $E_{\rm cen}$, $\delta$, and $B(M1)_{\Sigma}$ are from \cite{Pietr} and
references therein.}
\begin{tabular}{c|c|c|c|c|c|c|c|c|c}
\hline
   & &
\multicolumn{4}{|c|}{ $E_{\rm cen}$, MeV}    &
\multicolumn{4}{|c}{ $B(M1)_{\Sigma},\ \mu_N^2$}    \\
\cline{3-10}
 Nuclei & $\delta$ & exp & present & \cite{Urban} & $\Delta=0$ 
                   & exp & present & \cite{Urban} & $\Delta=0$ \\
\hline
 $^{150}$Nd & 0.22 & 3.04 & 2.88 & 3.44 & 1.92 & 1.61 & 1.64 & 4.17 & 7.26 \\[-.5mm]
 $^{152}$Sm & 0.24 & 2.99 & 2.99 & 3.46 & 2.02 & 2.26 & 2.50 & 4.68 & 7.81 \\[-.5mm]
 $^{154}$Sm & 0.26 & 3.20 & 3.10 & 3.57 & 2.17 & 2.18 & 3.34 & 5.42 & 8.65 \\[-.5mm]
 $^{156}$Gd & 0.26 & 3.06 & 3.09 & 3.60 & 2.16 & 2.73 & 3.44 & 5.42 & 8.76 \\[-.5mm]
 $^{158}$Gd & 0.26 & 3.14 & 3.09 & 3.60 & 2.19 & 3.39 & 3.52 & 5.72 & 9.12 \\[-.5mm]
 $^{160}$Gd & 0.27 & 3.18 & 3.14 & 3.61 & 2.21 & 2.97 & 4.02 & 5.90 & 9.38 \\[-.5mm]
 $^{160}$Dy & 0.26 & 2.87 & 3.08 & 3.59 & 2.13 & 2.42 & 3.60 & 5.53 & 9.03 \\[-.5mm]
 $^{162}$Dy & 0.26 & 2.96 & 3.07 & 3.61 & 2.14 & 2.49 & 3.69 & 5.66 & 9.25 \\[-.5mm]
 $^{164}$Dy & 0.26 & 3.14 & 3.07 & 3.60 & 2.17 & 3.18 & 3.78 & 5.95 & 9.59 \\[-.5mm]
 $^{164}$Er & 0.25 & 2.90 & 3.01 & 3.57 & 2.10 & 1.45 & 3.39 & 5.62 & 9.26 \\[-.5mm]
 $^{166}$Er & 0.26 & 2.96 & 3.06 & 3.53 & 2.13 & 2.67 & 3.86 & 5.96 & 9.59 \\[-.5mm]
 $^{168}$Er & 0.26 & 3.21 & 3.06 & 3.53 & 2.10 & 2.82 & 3.95 & 5.95 & 9.67 \\[-.5mm]
 $^{170}$Er & 0.26 & 3.22 & 3.05 & 3.57 & 2.09 & 2.63 & 4.03 & 5.91 & 9.79 \\[-.5mm]
 $^{172}$Yb & 0.25 & 3.03 & 2.99 & 3.55 & 2.05 & 1.94 & 3.72 & 5.84 & 9.79 \\[-.5mm]
 $^{174}$Yb & 0.25 & 3.15 & 2.98 & 3.47 & 2.02 & 2.70 & 3.80 & 5.89 & 9.82 \\[-.5mm]
 $^{176}$Yb & 0.24 & 2.96 & 2.92 & 3.45 & 1.94 & 2.66 & 3.46 & 5.54 & 9.58 \\[-.5mm]
 $^{178}$Hf & 0.22 & 3.11 & 2.81 & 3.43 & 1.79 & 2.04 & 2.67 & 4.86 & 9.00 \\[-.5mm]
 $^{180}$Hf & 0.22 & 2.95 & 2.81 & 3.36 & 1.76 & 1.61 & 2.69 & 4.85 & 8.97 \\[-.5mm]
 $^{182}$W  & 0.20 & 3.10 & 3.28 & 3.30 & 1.63 & 1.65 & 2.05 & 4.31 & 8.43 \\[-.5mm]
 $^{184}$W  & 0.19 & 3.31 & 3.24 & 3.28 & 1.55 & 1.12 & 1.72 & 3.97 & 8.14 \\[-.5mm]
 $^{186}$W  & 0.18 & 3.20 & 3.19 & 3.26 & 1.49 & 0.82 & 1.40 & 3.76 & 7.95 \\
\hline
\end{tabular}
\end{table}

\begin{table}\label{tab3a}
\caption{Scissors modes energy centroid $E_{\rm cen}$ and summarized
transition probabilities $B(M1)_{\Sigma}$. Parameters: $\kappa=0.05$ 
($\kappa=0.0637$ for $^{182, 184, 186}$W), $V_0=27$.}
\begin{tabular}{c|c|c|c|c|c|c|c|c|c}
\hline
   & &
\multicolumn{4}{|c|}{ $E_{\rm cen}$, MeV}    &
\multicolumn{4}{|c}{ $B(M1)_{\Sigma},\ \mu_N^2$}    \\
\cline{3-10}
 Nuclei & $\delta$ & exp & present & \cite{Urban} & $\Delta=0$ 
                   & exp & present & \cite{Urban} & $\Delta=0$ \\
\hline
 $^{134}$Ba & 0.14 & 2.99 & 3.04 & 3.09 & 1.28 & 0.56 & 0.68 & 1.67 & 3.90 \\[-.5mm]
 $^{148}$Nd & 0.17 & 3.37 & 3.22 & 3.18 & 1.48 & 0.78 & 1.28 & 2.58 & 5.39 \\[-.5mm]
 $^{150}$Sm & 0.16 & 3.13 & 3.17 & 3.13 & 1.42 & 0.92 & 1.12 & 2.45 & 5.26 \\[-.5mm]
 $^{182}$W  & 0.20 & 3.10 & 3.28 & 3.30 & 1.63 & 1.65 & 2.05 & 4.31 & 8.43 \\[-.5mm]
 $^{184}$W  & 0.19 & 3.31 & 3.24 & 3.28 & 1.55 & 1.12 & 1.72 & 3.97 & 8.14 \\[-.5mm]
 $^{186}$W  & 0.18 & 3.20 & 3.19 & 3.26 & 1.49 & 0.82 & 1.40 & 3.76 & 7.95 \\[-.5mm]
 $^{190}$Os & 0.15 & 2.90 & 3.14 & 3.12 & 1.21 & 0.98 & 1.38 & 2.67 & 6.64 \\[-.5mm]
 $^{192}$Os & 0.14 & 3.01 & 3.11 & 3.12 & 1.15 & 1.04 & 1.00 & 2.42 & 6.37 \\
\hline
\end{tabular}
\end{table}

The results of calculations for two groups ("light" and "heavy") of weakly deformed nuclei
with deformations $0.14\leq \delta \leq 0.17$ are shown in the Table 4.
They require some discussion, because of the 
self-consistency problem. These two groups of nuclei are transitional between well deformed
and spherical nuclei. Systematic calculations of equilibrium deformations \cite{Solov} 
predict $\delta_{eq}^{th}=0.0$ for $^{134}$Ba, $\pm 0.1$ for $^{148}$Nd,
 0.15 or -0.12 for $^{150}$Sm, 0.1 or -0.14 for $^{190}$Os and -0.1 for $^{192}$Os,
whereas their experimental values are $\delta_{eq}=0.14,\, 0.17,\, 0.16,\, 0.15$ and 0.14
respectively. As one sees, the discrepancy between theoretical and experimental 
$\delta_{eq}$ is large. Uncertain signs of theoretical equilibrium deformations are 
connected with very small ($\sim $0.1-0.2 MeV) difference between the values of deformation 
energies $\E_{def}$ at positive and negative $\delta_{eq}$. 
Even more so, the values of deformation
energies of these nuclei are very small: $\E_{def}=0.20,~0.50,~0.80$ and 0.70 MeV for
$^{148}$Nd, $^{150}$Sm, $^{190}$Os and $^{192}$Os respectively. This means that these
nuclei are very "soft" with respect of $\beta$- or $\gamma$-vibrations and probably they have
more complicated equilibrium shapes, for example,  hexadecapole or octupole deformations in
addition to the quadrupole one. This means that for the correct description of their dynamical
and equilibrium properties it is necessary to include higher order Wigner function moments
(at least fourth order) in addition to the second order ones. In this case it would be natural
also to use more complicate mean field potentials (for example, the Woods-Saxon one or the potential extracted 
from some of the numerous variants of Skyrme forces) instead of the too simple Nilsson potential.
Naturally, this will be the subject of further investigations. However, to be sure that the
situation with these nuclei is not absolutely hopeless, one can try
to imitate the properties of the more perfect potential by fitting
parameters of the Nilsson potential. As a matter of fact this potential has the single but
essential parameter -- the spin-orbital strength $\kappa_{Nils}$.
It turns out that changing its value from 0.0637 to 0.05 (the value used by Nilsson in his 
original paper \cite{Nils}) is enough to obtain the reasonable description of B(M1) factors
(see Table 4).
To obtain the reasonable description of the scissors energies we use the "freedom" of
choosing the value of the pairing interaction constant $V_0$ in (\ref{v_p}). It turns out
that changing its value from 25 MeV to 27 MeV is enough to obtain the satisfactory agreement
between the theoretical and experimental values of $E_{sc}$ (Table 4).

 The isotopes
$^{182-186}$W turn out intermediate between weakly deformed and well deformed nuclei: 
reasonable results  are obtained with $\kappa_{Nils}=0.0637$ (as for well deformed)
and $V_0=27$ MeV (as for weakly deformed). That is why they appear in both Tables.
 
Returning to the group of well deformed nuclei with $\delta \geq 0.18$ (Table 3) 
it is necessary to emphasize that all presented results for these nuclei were obtained without any 
fitting. In spite of it the agreement between the theory and experiment can be called 
excellent for all nuclei of this group except two: $^{164}$Er and $^{172}$Yb, where the 
theory overestimates B(M1) values approximately two times. However, these two nuclei fall out of the systematics and one can suspect, that
there the experimental B(M1) values are underestimated. Therefore one
can hope, that new experiments will correct the situation with these
nuclei, as it happened, for example, with $^{232}$Th \cite{Adekola}.

\section{Conclusion}

The method of Wigner function moments is generalized to take into account spin degrees of 
freedom and pair correlations simultaneously. 
The inclusion of the spin into the theory
allows one to discover several new phenomena. One of them, the nuclear spin scissors, was
described and studied in \cite{BaMo,BaMoPRC}, where some indications on the experimental
confirmation of its existence in actinides nuclei are discussed. Another phenomenon, the 
opposite rotation of spin up/down nucleons, or in other words, the phenomenon of hidden 
angular momenta, is described in this paper. Being determined by the spin degrees 
of freedom this phenomenon has  great influence on the excitation probability of the
spin scissors mode. On the other hand the spin scissors B(M1) values and the energies
of both, spin and orbital, scissors are very sensitive to the action of pair correlations.
As a result, these two factors, the spin up/down counter-rotation and pairing, working together, 
improve substantially the agreement between the theory and experiment in the description 
of the energy centroid of two nuclear scissors and their summed excitation probability.
More precisely, for the  first time an excellent agreement is achieved for well deformed 
nuclei of the rare earth region with standard values of all possible parameters.   

An  excellent agreement is also achieved for weakly deformed (transitional)
nuclei of the same region by a very modest re-fit of the spin-orbit strength.
We suppose that fourth order moments and more realistic interactions are required for the 
adequate description of transitional nuclei. However this shall be the subject of future work.

\vspace*{5mm}\noindent{\bf Acknowledgements}\\
Valuable discussions with V. N. Kondratyev are gratefully acknowledged.

\appendix

\renewcommand{\theequation}{A.\arabic{equation}}  
\setcounter{equation}{0} 
\section*{Appendix A}

{\bf Abnormal density}

According to formula (D.47) of \cite{Ring} the abnormal density in 
coordinate representation $\kappa(\br,s;\br',s')$ is connected with 
the abnormal density in the representation of the
harmonic oscillator quantum numbers
$\kappa_{\nu,\nu'}=\langle\Phi|a_{\nu}a_{\nu'}|\Phi\rangle$ by the relation
\begin{equation}
\kappa(\br,s;\br',s')=\langle\Phi|a(\br,s)a(\br',s')|\Phi\rangle=\sum_{\nu,\nu'}
\psi_{\nu}(\br,s)\psi_{\nu'}(\br',s')\langle\Phi|a_{\nu}a_{\nu'}|\Phi\rangle,
\label{kap1}
\end{equation}
where
$$\nu\equiv k,\varsigma,\quad k\equiv n,l,j,|m|, \quad \varsigma={\rm sign}(m)=\pm, \quad
k,+\equiv \nu, \quad k,-\equiv \bar \nu,
$$
$$\psi_{\bar \nu}(\br,s)=T\psi_{\nu}(\br,s).$$
$T$ - time reversal operator defined by formula (XV.85) of \cite{Mess}
$$T=-i\sigma_yK_0,$$
where $\sigma_y$ is the Pauli matrix and $K_0$ is the complex-conjugation
operator.

According to formula (7.12) of \cite{Ring}
$$a_{k,\varsigma}=u_k\alpha_{k,\varsigma}-\varsigma v_k\alpha^{\dagger}_{k,-\varsigma},$$
$$\alpha_{\nu}|\Phi\rangle=0,
$$
\begin{equation}
\langle\Phi|a_{\nu}a_{\nu'}|\Phi\rangle\equiv\kappa_{\nu\nu'}=-\varsigma'u_kv_{k'}
\langle\Phi|\alpha_{k,\varsigma}\alpha^{\dagger}_{k',-\varsigma'}|\Phi\rangle
=-\varsigma'u_kv_{k'}\delta_{k,k'}\delta_{-\varsigma,\varsigma'}.
\label{kapnn}
\end{equation}
This result means that in accordance with the theorem of Bloch and 
Messiah we have found the basis $|\nu\rangle$ in which the abnormal 
density $\kappa_{\nu,\nu'}$ has the canonical form. Therefore the spin
structure of $\kappa_{\nu,\nu'}$ is
\begin{equation}
\kappa_{\nu,\nu'}={\qquad 0\;\qquad u_{k}v_{k} 
\choose -u_{k}v_{k}\;\qquad 0 \qquad},
\label{anom}
 \end{equation}
or $\kappa_{\bar\nu,\nu}=-\kappa_{\nu,\bar\nu}$ and $\kappa_{\nu,\nu}=\kappa_{\bar\nu,\bar\nu}=0.$

With the help of (\ref{kapnn}) formula (\ref{kap1}) can be transformed into
\begin{eqnarray}
\kappa(\br,s;\br',s')
&=&\sum_{k,\varsigma}\varsigma u_kv_k
\psi_{k,\varsigma}(\br,s)\psi_{k,-\varsigma}(\br',s')
\nonumber\\
&=&\sum_{\nu>0}u_{\nu}v_{\nu}
[\psi_{\nu}(\br,s)\psi_{\bar \nu}(\br',s')
-\psi_{\bar \nu}(\br,s)\psi_{\nu}(\br',s')].
\label{kap2}
\end{eqnarray}
that reproduces formula (D.48) of \cite{Ring}.

{\bf What is the spin structure of $\kappa(\br,s;\br',s')$?}

Let us consider the spherical case:
$$\psi_{\nu}(\br,s)=\R_{nlj}(r)
\sum_{\Lambda,\sigma}C^{j m}_{l\Lambda,\frac{1}{2}\sigma}
Y_{l\Lambda}(\theta,\phi)\chi_{\frac{1}{2}\sigma}(s)
\equiv \R_{nlj}(r)\phi_{ljm}(\Omega,s),
$$
where $\phi_{ljm}(\Omega,s)=
\sum_{\Lambda,\sigma}C^{j m}_{l\Lambda,\frac{1}{2}\sigma}
Y_{l\Lambda}(\theta,\phi)\chi_{\frac{1}{2}\sigma}(s)$,
 spin function $\chi_{\frac{1}{2}\sigma}(s)=\delta_{\sigma,s}$
 and angular variables are denoted by $\Omega$.

Time reversal:
$$TY_{l\Lambda}=Y^*_{l\Lambda}=(-1)^{\Lambda}Y_{l-\Lambda},$$
$$T\chi_{\frac{1}{2}\frac{1}{2}}=\chi_{\frac{1}{2}-\frac{1}{2}},\quad
T\chi_{\frac{1}{2}-\frac{1}{2}}=-\chi_{\frac{1}{2}\frac{1}{2}}\quad
\to\quad T\chi_{\frac{1}{2}\sigma}
=(-1)^{\sigma-\frac{1}{2}}\chi_{\frac{1}{2}-\sigma},$$
$$T\sum_{\Lambda,\sigma}C^{jm}_{l\Lambda,\frac{1}{2}\sigma}Y_{l\Lambda}
\chi_{\frac{1}{2}\sigma}=
\sum_{\Lambda,\sigma}C^{jm}_{l\Lambda,\frac{1}{2}\sigma}Y_{l-\Lambda}
\chi_{\frac{1}{2}-\sigma}(-1)^{\Lambda+\sigma-\frac{1}{2}}=
\sum_{\Lambda,\sigma}C^{jm}_{l-\Lambda,\frac{1}{2}-\sigma}Y_{l\Lambda}
\chi_{\frac{1}{2}\sigma}(-1)^{-\Lambda-\sigma-\frac{1}{2}}$$
$$=\sum_{\Lambda,\sigma}C^{j-m}_{l\Lambda,\frac{1}{2}\sigma}Y_{l\Lambda}
\chi_{\frac{1}{2}\sigma}(-1)^{l+\frac{1}{2}-j-\Lambda-\sigma-\frac{1}{2}}=
\sum_{\Lambda,\sigma}C^{j-m}_{l\Lambda,\frac{1}{2}\sigma}Y_{l\Lambda}
\chi_{\frac{1}{2}\sigma}(-1)^{l-j+m}.$$
As a result
\begin{equation}
\psi_{\bar \nu}(\br,s)
=(-1)^{l-j+m}\R_{nlj}(r)\sum_{\Lambda,\sigma}
C^{j -m}_{l\Lambda,\frac{1}{2}\sigma}
Y_{l\Lambda}(\theta,\phi)\chi_{\frac{1}{2}\sigma}(s)
=(-1)^{l-j+m}\R_{nlj}(r)\phi_{lj-m}(\Omega,s),
\label{bark}
\end{equation}
that coincides with formula (2.45) of \cite{Ring}. Formula (\ref{kap2})
 can be rewritten now as
\begin{eqnarray}
\kappa(\br_1,s_1;\br_2,s_2)=\sum_{nljm>0}(uv)_{nljm}
\R_{nlj}(r_1)\R_{nlj}(r_2)(-1)^{l-j+m}
\nonumber\\
\left[
\phi_{ljm}(\Omega_1,s_1)\phi_{lj-m}(\Omega_2,s_2)-
\phi_{ljm}(\Omega_2,s_2)\phi_{lj-m}(\Omega_1,s_1)
\right]
\nonumber\\
=\sum_{nljm>0}(uv)_{nljm}
\R_{nlj}(r_1)\R_{nlj}(r_2)(-1)^{l-j+m}\sum_{\Lambda,\Lambda'}
\nonumber\\
\left[
C^{j m}_{l\Lambda,\frac{1}{2}s_1}C^{j -m}_{l\Lambda',\frac{1}{2}s_2}
Y_{l\Lambda}(\Omega_1)Y_{l\Lambda'}(\Omega_2)
-C^{j m}_{l\Lambda,\frac{1}{2}s_2}C^{j -m}_{l\Lambda',\frac{1}{2}s_1}
Y_{l\Lambda}(\Omega_2)Y_{l\Lambda'}(\Omega_1)
\right]
\nonumber\\
=\sum_{nljm>0}(uv)_{nljm}
\R_{nlj}(r_1)\R_{nlj}(r_2)(-1)^{l-j+m}\sum_{\Lambda,\Lambda'}
Y_{l\Lambda}(\Omega_1)Y_{l\Lambda'}(\Omega_2)
\nonumber\\
\left[
C^{j m}_{l\Lambda,\frac{1}{2}s_1}C^{j -m}_{l\Lambda',\frac{1}{2}s_2}
-C^{j m}_{l\Lambda',\frac{1}{2}s_2}C^{j -m}_{l\Lambda,\frac{1}{2}s_1}
\right].
\label{Rkap}
\end{eqnarray}
It is obvious that
$\kappa(\br,\uparrow;\br',\downarrow)\neq -\kappa(\br,\downarrow;\br',\uparrow)$, 
i.e. in the coordinate
representation the spin structure of $\kappa$ has nothing common with
(\ref{anom}).

The anomalous density defined by (\ref{Rkap}) has not definite angular
momentum $J$ and spin $S$. It can be represented as the sum of several 
terms with definite $J, S$. We have:
\begin{eqnarray}
\phi_{ljm}(1)\phi_{lj-m}(2)=
\sum_{0\le J \le 2j}C^{J0}_{jm,j-m}
\{\phi_j(1)\otimes\phi_j(2)\}_{J0}
\nonumber\\
=C^{00}_{jm,j-m}\{\phi_j(1)\otimes\phi_j(2)\}_{00}+
\sum_{1\le J \le 2j}C^{J0}_{jm,j-m}
\{\phi_j(1)\otimes\phi_j(2)\}_{J0}.
\label{phi12}
\end{eqnarray}
We are interested in the monopole pairing only, so we omit all terms 
except the first one:
\begin{eqnarray}
\left[\phi_{ljm}(1)\phi_{lj-m}(2)\right]_{J=0}=
C^{00}_{jm,j-m}\{\phi_j(1)\otimes\phi_j(2)\}_{00}
\nonumber\\
=(-1)^{j-m}\frac{1}{\sqrt{2j+1}}
\sum_{\nu,\sigma}C^{00}_{j\nu,j\sigma}
\phi_{j\nu}(1)\phi_{j\sigma}(2)
\nonumber\\
=\frac{1}{2j+1}\sum_{\nu}(-1)^{\nu-m}
\phi_{j\nu}(1)\phi_{j-\nu}(2).
\label{phiJ0}
\end{eqnarray}
Remembering the definition of $\phi$ function we find
\begin{eqnarray}
(-1)^m\left[\phi_{ljm}(\Omega_1,s_1)\phi_{lj-m}(\Omega_2,s_2)\right]_{J=0}
=\frac{1}{2j+1}\sum_{\nu}(-1)^{\nu}
\nonumber\\
\sum_{\Lambda,\sigma}
\sum_{\Lambda',\sigma'}
C^{j \nu}_{l\Lambda,\frac{1}{2}\sigma}
C^{j-\nu}_{l\Lambda',\frac{1}{2}\sigma'}
Y_{l\Lambda}(\Omega_1)
Y_{l\Lambda'}(\Omega_2)
\chi_{\frac{1}{2}\sigma}(s_1)
\chi_{\frac{1}{2}\sigma'}(s_2).
\label{phiJ0'}
\end{eqnarray}
The direct product of spin functions in this formula can be written as
\begin{eqnarray}
\chi_{\frac{1}{2}\sigma}(s_1)\chi_{\frac{1}{2}\sigma'}(s_2)=
\sum_{S,\Sigma}C^{S\Sigma}_{\frac{1}{2}\sigma,\frac{1}{2}\sigma'}
\{\chi_{\frac{1}{2}}(s_1)\otimes\chi_{\frac{1}{2}}(s_2)\}_{S\Sigma}
\nonumber\\
=C^{00}_{\frac{1}{2}\sigma,\frac{1}{2}\sigma'}
\{\chi_{\frac{1}{2}}(s_1)\otimes\chi_{\frac{1}{2}}(s_2)\}_{00}
+\sum_{\Sigma}C^{1\Sigma}_{\frac{1}{2}\sigma,\frac{1}{2}\sigma'}
\{\chi_{\frac{1}{2}}(s_1)\otimes\chi_{\frac{1}{2}}(s_2)\}_{1\Sigma}.
\label{chi12}
\end{eqnarray}
According to this result the formula for $\kappa$ consists of two
terms: the one with $S=0$ and another one with $S=1$.
It was shown in the paper \cite{Sandu} that the term 
with $S=1$ is an order of magnitude less than the term with $S=0$, so
we can neglect by it. Then
\begin{eqnarray}
\chi_{\frac{1}{2}\sigma}(s_1)\chi_{\frac{1}{2}\sigma'}(s_2)=
(-1)^{\frac{1}{2}-\sigma}\frac{1}{\sqrt2}\delta_{\sigma,-\sigma'}
\{\chi_{\frac{1}{2}}(s_1)\otimes\chi_{\frac{1}{2}}(s_2)\}_{00}
\nonumber\\
=(-1)^{\frac{1}{2}-\sigma}\frac{1}{\sqrt2}\delta_{\sigma,-\sigma'}
\sum_{\nu,\nu'}C^{00}_{\frac{1}{2}\nu,\frac{1}{2}\nu'}
\chi_{\frac{1}{2}\nu}(s_1)\chi_{\frac{1}{2}\nu'}(s_2)
\nonumber\\
=(-1)^{\frac{1}{2}-\sigma}\frac{1}{\sqrt2}\delta_{\sigma,-\sigma'}
\sum_{\nu=-1/2}^{1/2}(-1)^{\frac{1}{2}-\nu}\frac{1}{\sqrt2}
\chi_{\frac{1}{2}\nu}(s_1)\chi_{\frac{1}{2}-\nu}(s_2)
\nonumber\\
=(-1)^{\frac{1}{2}-\sigma}\frac{1}{2}\delta_{\sigma,-\sigma'}
\left[
\chi_{\frac{1}{2}\frac{1}{2}}(s_1)\chi_{\frac{1}{2}-\frac{1}{2}}(s_2)
-\chi_{\frac{1}{2}-\frac{1}{2}}(s_1)\chi_{\frac{1}{2}\frac{1}{2}}(s_2)
\right]
\nonumber\\
=\frac{1}{2}\delta_{\sigma,-\sigma'}(-1)^{\frac{1}{2}-\sigma}
\left[
\delta_{s_1 \frac{1}{2}}\delta_{s_2 -\frac{1}{2}}
-\delta_{s_1 -\frac{1}{2}}\delta_{s_2 \frac{1}{2}}
\right].
\label{chiS0}
\end{eqnarray}
Inserting this result into (\ref{phiJ0'}) we find
\begin{eqnarray}
(-1)^m\left[\phi_{ljm}(\Omega_1,s_1)\phi_{lj-m}(\Omega_2,s_2)\right]_{J=0}^{S=0}
=\frac{1}{2}
\left[
\delta_{s_1 \frac{1}{2}}\delta_{s_2 -\frac{1}{2}}
-\delta_{s_1 -\frac{1}{2}}\delta_{s_2 \frac{1}{2}}
\right]
\frac{1}{2j+1}
\nonumber\\
\sum_{\Lambda,\Lambda'}
Y_{l\Lambda}(\Omega_1)Y_{l\Lambda'}(\Omega_2)
\sum_{\nu,\sigma}(-1)^{\nu+\frac{1}{2}-\sigma}
C^{j \nu}_{l\Lambda,\frac{1}{2}\sigma}
C^{j-\nu}_{l\Lambda',\frac{1}{2}-\sigma}
\nonumber\\
=\frac{1}{2}
\left[
\delta_{s_1 \frac{1}{2}}\delta_{s_2 -\frac{1}{2}}
-\delta_{s_1 -\frac{1}{2}}\delta_{s_2 \frac{1}{2}}
\right]
\frac{1}{2j+1}
\nonumber\\
\sum_{\Lambda,\Lambda'}
Y_{l\Lambda}(\Omega_1)Y_{l\Lambda'}(\Omega_2)
\sum_{\nu,\sigma}
(-1)^{\frac{1}{2}+\Lambda}
\frac{2j+1}{2l+1}(-1)^{1+j+\frac{1}{2}-l}
C^{l\Lambda}_{j\nu,\frac{1}{2}-\sigma}
C^{l-\Lambda'}_{j\nu,\frac{1}{2}-\sigma}
\nonumber\\
=\frac{1}{2}
\left[
\delta_{s_1 \frac{1}{2}}\delta_{s_2 -\frac{1}{2}}
-\delta_{s_1 -\frac{1}{2}}\delta_{s_2 \frac{1}{2}}
\right]
\frac{1}{2l+1}
(-1)^{j-l}\sum_{\Lambda,\Lambda'}
Y_{l\Lambda}(\Omega_1)Y_{l\Lambda'}(\Omega_2)
(-1)^{\Lambda}\delta_{\Lambda,-\Lambda'}
\nonumber\\
=\frac{1}{2}
\left[
\delta_{s_1 \frac{1}{2}}\delta_{s_2 -\frac{1}{2}}
-\delta_{s_1 -\frac{1}{2}}\delta_{s_2 \frac{1}{2}}
\right]
(-1)^{j-l}\frac{1}{4\pi}
P_{l}(\cos\Omega_{12}),
\label{phiJS0}
\end{eqnarray}
where $P_{l}(\cos\Omega_{12})$ is Legendre polynomial and $\Omega_{12}$ 
is the angle between vectors $\br_1$ and $\br_2$.
With the help of this result formula (\ref{Rkap}) is transformed into
\begin{eqnarray}
\kappa(\br_1,s_1;\br_2,s_2)_{J=0}^{S=0}=
\left[
\delta_{s_1 \frac{1}{2}}\delta_{s_2 -\frac{1}{2}}
-\delta_{s_1 -\frac{1}{2}}\delta_{s_2 \frac{1}{2}}
\right]
\frac{1}{4\pi}\sum_{nljm>0}(uv)_{nljm}
\R_{nlj}(r_1)\R_{nlj}(r_2)P_{l}(\cos\Omega_{12}).
\label{kapJS0}
\end{eqnarray}
Now it is obvious that in the coordinate representation $\kappa$ with
$J=0, S=0$ has the spin structure similar to the one demonstrated by
formula (\ref{anom}):
\begin{equation}
\kappa(\br_1,s_1;\br_2,s_2)_{J=0}^{S=0}=
{\qquad 0\;\qquad \kappa(\br_1,\br_2)
\choose -\kappa(\br_1,\br_2)\;\qquad 0 \qquad}
\label{Ranom}
\end{equation}
with
\begin{eqnarray}
\kappa(\br_1,\br_2)=
\frac{1}{4\pi}\sum_{nljm>0}(uv)_{nljm}
\R_{nlj}(r_1)\R_{nlj}(r_2)P_{l}(\cos\Omega_{12}).
\label{kap0}
\end{eqnarray}

\renewcommand{\theequation}{B.\arabic{equation}}  
\setcounter{equation}{0} 
\section*{Appendix B}

{\bf Wigner transformation}

The Wigner Transform (WT) of the single-particle operator matrix
$\hat F_{\br_1,\sigma;\br_2,\sigma'}$ is defined as
\begin{eqnarray}
[\hat F_{\br_1,\sigma;\br_2,\sigma'}]_{\rm WT}\equiv
F_{\sigma,\sigma'}(\br,\bp)
=\int d^3s e^{-i\bp\cdot\bs/\hbar}
\hat F_{\br+\bs/2,\sigma;\br-\bs/2,\sigma'}
\end{eqnarray}
with $\br=(\br_1+\br_2)/2$ and $\bs=\br_1-\br_2.$
It is easy to derive a pair of useful relations. The first one is
\begin{eqnarray}
F_{\sigma,\sigma'}^*(\br,\bp)\!\!\!\!\!
&&=\int d^3s e^{i\bp\cdot\bs/\hbar}
\hat F^*_{\br+\bs/2,\sigma;\br-\bs/2,\sigma'}
=\int d^3s e^{-i\bp\cdot\bs/\hbar}
\hat F^*_{\br-\bs/2,\sigma;\br+\bs/2,\sigma'}
\nonumber\\
&&=\int d^3s e^{-i\bp\cdot\bs/\hbar}
\hat F^{\dagger}_{\br+\bs/2,\sigma';\br-\bs/2,\sigma}=
[\hat F^{\dagger}_{\br_1,\sigma';\br_2,\sigma}]_{\rm WT},
\end{eqnarray}
i.e., $[\hat F^{\dagger}_{\br_1,\sigma;\br_2,\sigma'}]_{\rm WT}
=[\hat F_{\br_1,\sigma';\br_2,\sigma}]_{\rm WT}^*
=F_{\sigma'\sigma}^*(\br,\bp).$
The second relation is
\begin{eqnarray}
\bar F_{\sigma\sigma'}(\br,\bp)\!\!\!\!\!&&\equiv F_{\sigma\sigma'}(\br,-\bp)
=\int d^3s e^{i\bp\cdot\bs/\hbar}
\hat F_{\br+\bs/2,\sigma;\br-\bs/2,\sigma'}
\nonumber\\
&&=\int d^3s e^{-i\bp\cdot\bs/\hbar}
\hat F_{\br-\frac{\bs}{2},\sigma;\br+\frac{\bs}{2},\sigma'}
=\int d^3s e^{-i\bp\cdot\bs/\hbar}
[\hat F^{\dagger}_{\br+\bs/2,\sigma';\br-\bs/2,\sigma}]^*.
\end{eqnarray}
For the hermitian operators $\hat \rho$ and $\hat h$ this latter relation gives
$[\hat\rho^*_{\br_1,\sigma;\br_2,\sigma}]_{\rm WT}
=\rho_{\sigma\sigma}(\br,-\bp)$ and
$[\hat h^*_{\br_1,\sigma;\br_2,\sigma}]_{\rm WT}
=h_{\sigma\sigma}(\br,-\bp)$.

The Wigner transform of the product of two matrices $F$ and $G$ is
\begin{equation}
[\hat F\hat G]_{\rm WT}=F(\br,\bp)\exp\left(\frac{i\hbar}{2}
\stackrel{\leftrightarrow}{\Lambda}\right)G(\br,\bp),
\end{equation}
where the symbol
 $\stackrel{\leftrightarrow}{\Lambda}$
 stands for the Poisson bracket operator
$$
\stackrel{\leftrightarrow}{\Lambda}
=\sum_{i=1}^3\left(
\frac{\stackrel{\gets}{\partial} }{\partial r_i}\frac{
\stackrel{\to}{\partial} }{\partial p_i}
-\frac{\stackrel{\gets}{\partial} }{\partial p_i}\frac{
\stackrel{\to}{\partial} }{\partial r_i}\right)\,.$$

\renewcommand{\theequation}{C.\arabic{equation}}  
\setcounter{equation}{0} 
\section*{Appendix C}

{\bf Integrals of motion}

 Isovector integrals of motion: 
\begin{eqnarray}
\label{intmot}
     &&i\hbar\frac{\eta}{2} {\L}^{+}_{21}
-\hbar^2\frac{\eta^2 m}{8}\left[\R^{-}_{21}+2\R^{\u}_{22}\right]
+\sqrt{\frac{2}{3}}\left(\frac{3}{8}\hbar^2\eta^2 m-c_3\right)\R^\d_{20}
+\sqrt{\frac{2}{3}}\frac{1}{m}\P^\d_{20}
\nonumber\\
&&+\frac{1}{2\sqrt3 c_2}\left((c_1-c_2)(c_1+2c_2)+2c_1c_3-\frac{3}{2}\hbar^2\eta^2 m\right)\R^\d_{00}
\nonumber\\
&&+\frac{1}{\sqrt3 c_2 m}\left(c_1+c_2+2c_3-\frac{3}{2}\hbar^2\eta^2 m\right)\P^\d_{00}
=const,
\nonumber\\
    && i\hbar\frac{\eta}{2} \left[\L^{+}_{11}-i\frac{\hbar}{2} F^\d\right]
-3\sqrt6(1-\alpha)\kappa_0 R_{20}^{\rm eq}
\left[
\frac{2}{\sqrt3 c_2 m}\P^\d_{00}+\frac{c_1}{\sqrt3 c_2}\R^\d_{00}-\sqrt{\frac{2}{3}}\R^\d_{20}
\right]=const,
\nonumber\\
 && i\hbar\frac{3}{4}\eta c_2\tilde\L_{11} + \frac{\Delta_0(r')}{\hbar}
 \left\{
 i\hbar\frac{\eta}{2}\left[\P^-_{21}+\frac{m}{4}(2c_1+c_2)\R^-_{21}-\sqrt{\frac{2}{3}}\P^\d_{20}\right]
\right.\nonumber\\
&& - \left(i\hbar\frac{\eta}{4\sqrt2} -\frac{4\sqrt6}{mc_2}\kappa_0\alpha L_{10}^{\rm eq}\right)\P^\d_{00}
   - \left(i\hbar\frac{\eta m}{2}\sqrt{\frac{2}{3}}(2c_1+c_2) +4\sqrt3\kappa_0\alpha L_{10}^{\rm eq}\right)\R^\d_{20}
\nonumber\\
&&\left.- \left(i\hbar\frac{\eta m}{8\sqrt3}(c_1-4c_2) - 2\sqrt2\frac{c_1}{c_2}\kappa_0\alpha L_{10}^{\rm eq}\right)\R^\d_{00}
\right\} = const,
\nonumber\\
    && \P^\u_{22} - \sqrt{\frac{2}{3}}\left(\P^\d_{20}+\sqrt2 \P^\d_{00}\right)+\frac{m}{2}(c_1-c_2)
    \left[\R^\u_{22} - \sqrt{\frac{2}{3}}\left(\R^\d_{20}+\sqrt2 \R^\d_{00}\right)\right] = const,
\nonumber\\
&& i\hbar\frac{\eta}{2} \tilde\R_{21}
 -\left( \frac{16}{5\hbar} \kappa_0\alpha\K_4+\frac{\Delta_0(r')}{\hbar}-\frac{3}{8}\hbar\chi\kappa_0(r')\! \right)\!
  \!\!\left[\sqrt{\frac{2}{3}}\R^\d_{20} -\frac{c_1}{\sqrt3 c_2} \R^\d_{00}- \frac{2}{\sqrt3 mc_2} \P^\d_{00}\right] = const,
\nonumber\\
&& i\hbar\frac{\eta}{2} \tilde\P_{21}
 -\frac{\Delta_0(r')}{\hbar}
\left[\sqrt{\frac{2}{3}}\P^\d_{20} + \frac{2(c_1+c_2)}{\sqrt3 c_2} \P^\d_{00}+\frac{m}{2}\frac{(c_1-c_2)(c_1+2c_2)}{\sqrt3 c_2} \R^\d_{00}
\right]
\nonumber\\
&& + \,6\hbar\kappa_0\alpha\K_0
 \left[\sqrt{\frac{2}{3}}\R^\d_{20} -\frac{c_1}{\sqrt3 c_2} \R^\d_{00}- \frac{2}{\sqrt3 mc_2} \P^\d_{00}\right] = const,
\nonumber\\
&& \tilde\L_{21}
 +\frac{\Delta_0(r')}{\hbar}
\left[\frac{1}{\sqrt3 c_2}\P^\d_{00} + \frac{m}{2} \left( \R^-_{21}-\sqrt{\frac{2}{3}}\R^\d_{20}+\frac{c_1}{\sqrt3 c_2}\R^\d_{00}\right)\right]
= const,
\end{eqnarray}
where
\begin{eqnarray}
&&c_1\equiv 2m\omega^2-
\frac{\sqrt3}{2}\hbar^2 \chi I_2 \frac{\left(2\A_1-\A_2\right)}{\A_1\A_2},
\qquad c_2\equiv 4\sqrt6\kappa_0 R_{20}^{\rm eq}+
\frac{\sqrt3}{2}\hbar^2 \chi I_2 \frac{\left(\A_1+\A_2\right)}{\A_1\A_2},
\nonumber\\
&&c_3\equiv m\,\omega^2-4\sqrt3\alpha\kappa_0R_{00}^{\rm eq}+\sqrt6(1+\alpha)\kappa_0 R_{20}^{\rm eq}.
\nonumber
\end{eqnarray}

Isoscalar integrals of motion are easily obtained from isovector ones by taking $\alpha=1$. In the case of harmonic oscillations 
all constants $const$ are obviously equal to zero.

\renewcommand{\theequation}{D.\arabic{equation}}  
\setcounter{equation}{0} 
\section*{Appendix D}

\begin{eqnarray}
&&I^{\kappa\Delta}_{pp}(\br,p)=
\frac{r_p^3}{\sqrt{\pi}\hbar^3}e^{-\alpha p^2}
\int\!\kappa^r(\br,p')\left[\phi_0(x)
-4\alpha^2p'^4\phi_2(x)\right]
e^{-\alpha p'^2}p'^2dp',
\\
&&I^{\kappa\Delta}_{rp}(\br,p)=
\frac{r_p^3}{\sqrt{\pi}\hbar^3}e^{-\alpha p^2}
\int\!\kappa^r(\br,p')[\phi_0(x)
-2\alpha p'^2\phi_1(x)]e^{-\alpha p'^2}p'^2dp',
\end{eqnarray}
where $x=2\alpha pp'$,

\begin{eqnarray}
&&\phi_0(x)=\frac{1}{x}\sinh(x),
\qquad\phi_1(x)=\frac{1}{x^2}\left[\cosh(x)-\frac{1}{x}\sinh(x)\right],
\nonumber \\
&&\phi_2(x)=\frac{1}{x^3}\left[\left(1
+\frac{3}{x^2}\right)\sinh(x)-\frac{3}{x}\cosh(x)\right].
\end{eqnarray}

Anomalous density and semiclassical gap equation \cite{Ring}:
\begin{eqnarray}
&&\kappa(\br,\bp)=\frac{1}{2}
\frac{\Delta(\br,\bp)}{\sqrt{h^2(\br,\bp)+\Delta^2(\br,\bp)}},
\\
&&\Delta(\br,\bp)=-\frac{1}{2}\int\!\frac{d^3p'}{(2\pi\hbar)^3}
v(|\bp-\bp'|)
\frac{\Delta(\br,\bp')}{\sqrt{h^2(\br,\bp')+\Delta^2(\br,\bp')}},
\end{eqnarray}
where $v(|\bp-\bp'|)=\beta e^{-\alpha |\bp-\bp'|^2}\!$
with $\beta=-|V_0|(r_p\sqrt{\pi})^3$ and $\alpha=r_p^2/4\hbar^2$.


\begin{thebibliography}{99}

\bibitem{Hilt}
R. R. Hilton "A possible vibrational mode in heavy nuclei",
Int. Conf. on Nuclear Structure (Dubna, June 1976), unpublished.

\bibitem{Hilt92}
R. R. Hilton, Ann. Phys. (N.Y.) 214 (1992) 258.

\bibitem{Suzuki}
 T. Suzuki, D. J. Rowe, Nucl. Phys. A 289 (1977) 461.

\bibitem{Lo}
N. Lo Iudice, F. Palumbo, Phys. Rev. Lett. 41 (1978) 1532.

\bibitem{Zaw}
D. Zawischa, J. Phys. G: Nucl. Part. Phys. 24 (1998) 683.

\bibitem{Sushkov}
 V. G. Soloviev, A. V. Sushkov, N. Yu. Shirikova and N. Lo Iudice,
 Nucl. Phys. A 600 (1996) 155.

\bibitem{Lo2000}
N. Lo Iudice, La Rivista del Nuovo Cimento 23 (2000) N.9.

\bibitem{Lipp}
E. Lipparini, S. Stringari, Phys. Rep. 175 (1989) 103.

\bibitem{Heyd} 
K. Heyde, P. von Neuman-Cosel and A. Richter,
Rev. Mod. Phys. 82 (2010) 2365.

\bibitem{Kneis}
 U. Kneissl, H. H. Pitz, and A. Zilges, Prog. Part. Nucl. Phys. 37 (1996) 349.

\bibitem{Richt}
 A. Richter, Prog. Part. Nucl. 34 (1995) 261.

\bibitem{Malov}
E. B. Balbutsev, L. A. Malov, P. Schuck, M. Urban, and X. Vi\~nas, Phys. At. Nucl. 71 (2008) 1012.

\bibitem{Urban}
E. B. Balbutsev, L. A. Malov, P. Schuck, and M. Urban, Phys. At. Nucl. 72 (2009) 1305.

\bibitem{BaMo}
E. B. Balbutsev, I.V. Molodtsova, P. Schuck, Nucl. Phys. A 872 (2011) 42.

\bibitem{BaMoPRC}
E. B. Balbutsev, I.V. Molodtsova, P. Schuck, Phys. Rev. C 88 (2013) 014306.

\bibitem{Solov}
V. G. Soloviev, {\it Theory of complex nuclei} (Pergamon Press, Oxford, 1976). 

\bibitem{Ring} P. Ring and P. Schuck,
 {\it The Nuclear Many-Body Problem} (Springer, Berlin, 1980).

\bibitem{M.Urban} M. Urban, Phys. Rev. A {\bf 75}, 053607 (2007).

\bibitem{Var}
 D. A. Varshalovitch, A. N. Moskalev and V. K. Khersonski,
{\it Quantum Theory of Angular Momentum} (World Scientific, Singapore, 1988).

\bibitem{BaSc}
E. B. Balbutsev, P. Schuck, Nucl. Phys. A 720 (2003) 293;\\
E. B. Balbutsev, P. Schuck, Nucl. Phys. A 728 (2003) 471.

\bibitem{Ann}
E. B. Balbutsev, P. Schuck, Ann. Phys. 322 (2007) 489.

\bibitem{BrMt}
A. Bohr, B. Mottelson, {\it Nuclear Structure, Vol. 2} (Benjamin, New York, 1975).

\bibitem{Moya}
P. Sarriguren, E. Moya de Guerra, R. Nojarov, Phys. Rev. C 54 (1996) 690;\\
P. Sarriguren, E. Moya de Guerra, R. Nojarov, Z. Phys. A 357  (1997) 143.

\bibitem{Giai}
N. Van Giai, H. Sagawa, Phys. Lett. B 106 (1981) 379.

\bibitem{Floc}
 M. Beiner, H.Flocard, N. Van Giai, P. Quentin, Nucl. Phys. A 238 (1975) 29.

\bibitem{Nils} 
S. G. Nilsson, Mat.-fys. Medd. Dan. Vid. Selsk. {\bf 29} (1955) 16.

\bibitem{Pietr}
N. Pietralla, P. von Brentano, R.-D. Herzberg, U. Kneissl, N. Lo Iudice, H. Maser,
H. H. Pitz, and A. Zilges, Phys. Rev. C {\bf 58}, 184 (1998).

\bibitem{Adekola}
A. S. Adekola, C. T. Angell, S. L. Hammond, A. Hill, C. R. Howell, H. J. Karwowski, 
J. H. Kelley, and E. Kwan, Phys. Rev. C {\bf 83} (2011) 034615.

\bibitem{Mess} A.Messiah,
 {\it Quantum Mechanics, Vol. 2} (North Holland, Amsterdam, 1961).

\bibitem{Sandu}
N. Pillet, N. Sandulescu, P. Schuck, J.-F. Berger, Phys. Rev. C 81 (2010) 034307.

\end{thebibliography}
\end{document}